\documentclass[reprint,aps,9pt]{revtex4-1}

\usepackage[english]{babel}

\usepackage{bm}    
\usepackage{esvect}
\usepackage{amsmath}    
\usepackage{graphicx}   
\usepackage{verbatim}   
\usepackage{color}      
\usepackage{subfigure}  
\usepackage{hyperref}   
\newcommand*{\refequa}[1]{(\ref{#1})}

\newcommand*{\partialD}[2] {\frac{\partial #1}{\partial #2}}

\renewcommand{\vec}[1]{\mathbf{#1}}
\newcommand{\beq}{\begin{equation}}
\newcommand{\eeq}{\end{equation}}
\newcommand{\tobs}{t_{\rm obs}}
\newcommand{\gammaobs}{\gamma_{\mathrm{obs}}}

\newcommand{\LL}{ L_{\rm r} }  

\begin{document}
	\title{Long-ranged correlations in large deviations of local clustering}
	\author{Jakub Dolezal$^1$, Robert L. Jack$^{1,2}$}
	\affiliation{$^1$ DAMTP, University of Cambridge, Centre for Mathematical
		Sciences, Wilberforce Road, Cambridge CB3 0WA, United Kingdom}

	\affiliation{$^2$ Department of Chemistry, University of Cambridge, Lensfield
		Road, Cambridge CB2 1EW, United Kingdom}

\begin{abstract}
		In systems of diffusing particles, we investigate large deviations of a time-averaged measure of clustering around one particle.  We focus on biased ensembles of trajectories, which realise large-deviation events.  The bias acts on a single particle, but elicits a response that spans the whole system.
		We analyse this effect through the lens of Macroscopic Fluctuation Theory, focussing on the coupling of the bias to hydrodynamic modes.  This explains that the dynamical free energy has non-trivial scaling relationships with the system size, in 1 and 2 spatial dimensions.  
		We show that the long-ranged response to a bias on one particle also has consequences when biasing two particles.
\end{abstract}

\maketitle

\section{Introduction}

Simple systems of interacting particles with diffusive dynamics exhibit a wealth of dynamical fluctuation 
behaviour~\cite{Bertini2005,Bertini2006,Bertini2015,PhysRevLett.119.090602,YongJooDynamicalSymmetryBreaking, BODINEAU2007540, Jack2010, Dhar_2014, Jack2015,Klymko2017}.
Of particular interest are \emph{large deviations}, which involve long-lived fluctuations~\cite{Touchette2009,Derrida2007,Jack2020ergo}, and
are often associated with collective behaviour~\cite{Nemoto2019,hurtado_thermodynamics_2014, Jack2015, Fodor_2020,Dolezal_2019,Jack2010,YongJooDynamicalSymmetryBreaking, Dhar_2014, Barato2015}.
 Large deviations play an important role in statistical physics, as a way to analyse the thermodynamic limit in equilibrium systems~\cite{Touchette2009}, and the convergence of time averages (ergodicity)~\cite{Touchette2009,Derrida2007,Jack2020ergo}
 
In the dynamical context, an important role is played by \emph{biased ensembles of trajectories}, which reveal the mechanism by which large deviations occur.  One introduces a conjugate field for the time-averaged quantity of interest, and studies the response to this field.  Singular responses can be interpreted as \emph{dynamical phase transitions}, which can be associated with spontaneous symmetry breaking~\cite{Jack2010,YongJooDynamicalSymmetryBreaking,Nemoto2019,keta2020collective} and/or long-ranged correlations~\cite{Bodineau2008,Jack2015,Dolezal_2019}.  An important class of biased ensembles applies to systems with diffusive dynamics, which can be analysed by macroscopic fluctuation theory (MFT)~\cite{Bertini2015}.  In these systems, the response is controlled by hydrodynamic modes, which can result in universal predictions that depend only on the diffusivity and mobility, independent of model details~\cite{appert-rolland_universal_2008, Dolezal_2019, lecomte_inactive_2012,Bertini2005, Bertini2006,Bertini2015}.

Biased ensembles have also been analysed for large deviations of the dynamical activity, which is particularly relevant in glassy systems~\cite{Hedges2009, Garrahan2007,Garrahan2009,speck_constrained_2012,speck_first-order_2012} as well as in diffusive systems~\cite{appert-rolland_universal_2008,Jack2015,Dolezal_2019}.  In the glassy context, a bias to low activity is associated with phase transitions into dynamically arrested states; in diffusive systems one more often observes transitions into inhomogeneous (or phase-separated) states~\cite{lecomte_inactive_2012,Dolezal_2019}.
These previous works have considered the \emph{total} dynamical activity in large systems, obtained by summing over all particles (or all sites on a lattice).  In this work we consider large deviations of an activity-like quantity that involves just one or two \emph{tagged particles}, in a large system.   
(Specifically, we consider a measure of local clustering in place of a single-particle activity, the relation between these quantities is explained below.)
In the associated biased ensembles, the field only couples to the tagged particle(s), leading to a localized bias.  We find that the response to this local bias is long-ranged, and spans the whole system.  

This result may be unexpected, particularly if one invokes the analogy between thermodynamic ensembles and ensembles of dynamical trajectories~\cite{Touchette2009, jack2005, Lecomte2007, Garrahan2009}, where one may expect a localized response to local biasing fields.  This work explains that the reason for this counterintuitive result is the coupling of the bias to hydrodynamic modes.  We analyse this coupling using MFT arguments, which we compare with numerical results obtained by transition path sampling (TPS)~\cite{Bolhuis, Hedges2009}.  Overall, these results further emphasise earlier insights that systems with hydrodynamic modes have characteristic responses in biased ensembles~\cite{Bertini2015,Derrida2007}, that may not be predicted based on analogies with thermodynamic ensembles~\cite{Jack2020ergo}.  These effects are particularly strong in low-dimensional systems, including $d=1$ and $d=2$. As a particular example, we show here that applying a weak bias to two tagged particles (in $d=1$) can cause them to localise near each other, even in a very large system.

To place our results in context, we recall some previous work on large-deviations of single-particle quantities~\cite{Imparato2009,Krapivsky2014,Dhar_2014,Cagnetta2017,Grandpre2018}.
In one dimension, exact results are available for large displacements of single tracer particles,  
which cannot overtake their neighbours because of the hard-core interaction~\cite{Imparato2009,Krapivsky2014,Dhar_2014}, and for how locally induced current affects a system\cite{Sadhu2011}.  Hence a large displacement of one particle necessarily involves a large displacement of its neighbours, leading to a long-ranged coupling.   By contrast, the fluctuations considered here do not require a large displacement of the tagged particle, only that its local environment should be non-typical.  We expand further on this distinction in later sections. 
Results for active systems~\cite{Cagnetta2017,Grandpre2018} indicate that large deviations for single particles are correlated with the behaviour of nearby particles, but the issue of long-ranged correlations has not been discussed in detail.  We show here that such correlations play an important role in the large-deviation behavior.

The article is organized as follows.  {In section \ref{sec::Model} we introduce the different models of Brownian particles, and we define our measures of clustering and the associated biased ensembles. In section \ref{sec::Results1D} we discuss the one-dimensional version of the model, including the variance of the clustering and its large deviations; we also compare the results with predictions from MFT.
In section \ref{sec::Results1D2p} we investigate the effects of biasing two particles at once.  Section \ref{sec:Results2D} discusses clustering around one particle in two dimensions. Finally in section \ref{sec::Summary} we discuss the implications and conclude.}

\section{Definitions}\label{sec::Model}

\subsection{Model}

We consider models of interacting particles in $d$ spatial dimensions, specifically $d=1$ and $d=2$.  The number of particles is $N$ and the
position vector of the $i$th particle is $\vec{x}_i$, which lies within a periodic box of volume $L^d$.  
The mean density is 
\begin{equation}
\rho_0 = \frac{ N }{ L^d } \; .
\end{equation}
The overdamped Langevin equation governing the movement of particle $i$ is 
\begin{equation}\label{equa::origLangevin}
		\dot{\vec{x}}_i = -\beta D_0 \nabla_i U + \sqrt{2D_0} \bm{\eta}_i \; ,
\end{equation}
where $\bm{\eta}_i$ is a Brownian white noise, $D_0$ is the (bare) diffusion constant of these particles, $\beta$ the inverse temperature, and $U$ the interaction energy, which is of the form
\begin{equation}
	U=\sum_{1\leq j < k \leq N} v(\lvert\vec{x}_j-\vec{x}_k\rvert) \; .
\end{equation}
 Here $v(r)$ is a short-ranged interaction potential. 

For $d=1$ we take a hard-core interaction such that the particles have size $l_0$:
\begin{equation}
\label{equ:pair-inf}
v(x)= \begin{cases} 
0,  & \text{$x>l_0$} \; , \\ 
\infty,  &  \text{$x<l_0$} \; .
\end{cases}
\end{equation}
The consistency of this potential with (\ref{equa::origLangevin}) is ensured by using a Monte Carlo (MC) dynamics to approximate the 
Langevin dynamics~\cite{gardiner2004handbookchap4}, as in \cite{Dolezal_2019}.  The maximal step size for the trial MC moves is $a$.
We take $a<l_0$ so particles cannot pass each other, and it is consistent to order their positions such that $x_1<x_2<\dots$, modulo periodic boundaries.  It is also useful to define a reduced system size 
\begin{equation}
\LL=L-Nl_0
\label{equ:LL}
\end{equation} 
and a corresponding set of positions $\tilde x_j = x_j-jl_0$ such that $\tilde x_1 < \tilde x_2 < \dots$ and every $\tilde x_j \in [0,\LL)$.  The relation between the reduced system and the original one is discussed in~\cite{Dolezal_2019}, a useful feature of the reduced system is that the Boltzmann distribution for its equilibrium state has particles distributed as an ideal gas (as long as the labels $j$ are ignored).

For $d=2$ we consider a Weeks-Chandler-Andersen
(WCA) potential \cite{Weeks1971} of strength $\epsilon$ and range $l_0$:
\begin{equation}
v(r)=
\begin{cases}
 4\epsilon \left[ (l_0/r)^{12} - (l_0/r)^6 \right] + \epsilon \; ,
& \text{$r< 2^{1/6}l_0$} \; ,
\\ 0 \; ,  & \text{$r>2^{1/6}l_0$} \; .
\end{cases}
\end{equation} 
In this case we integrate the equation of motion (\ref{equa::origLangevin}) using the Euler-Maruyama method~\cite{Kloeden1994}.
For both cases, the natural time unit is the Brownian time
\begin{equation}
\tau_{\rm B}=\frac{l^2_0}{2D_0} \; .
\end{equation}

We explain below (Sec.~\ref{sec:hydro-1}) that the behaviour discussed here depends weakly on the specific shape of the potential, and similar results would be expected for any system with sufficiently strong short-ranged repulsive forces.  For numerical work we set the unit of length by taking $l_0=1$, but for theoretical analysis we retain this quantity as an explicit parameter. Likewise we set $\beta=1$ in numerical work.

\subsection{Time-averaged clustering around a tagged particle}

As anticipated above, we focus on rare events where a single tagged particle behaves in a non-typical way.  Specifically, we define measurements of clustering, which depend on the local environment of the tagged particle, and are sensitive to whether the local density is higher or lower than the average.  Supposing that the tagged particle has index $i$, we denote the local clustering at time $t$ by $c_i(t)$.  Specific expressions are given in (\ref{equ:ci},\ref{equ:ci-2d}) below.  

Now define the observation time $\tobs$, and the time-averaged measure of clustering as
\begin{equation}
	c[\mathbf{x}]=\frac{1}{\tobs}\int_{0}^{\tobs}\; c_i(t)\mathrm{d}t
	\label{equ:cbar}
\end{equation}
where the notation $[\mathbf{x}]$ indicates that this quantity depends on the trajectory followed by the stochastic dynamics, over the time period $[0,\tobs]$.  As $\tobs\to\infty$, this (random) quantity obeys a large deviation principle (LDP) which means 
its probability density scales as 
\beq
p(c|\tobs,L) \sim {\rm e}^{-\tobs I_L(c) },
\label{equ:ldp}
\eeq
where $I_L(c)$ is the rate function.
(Note, the limit $\tobs\to\infty$ is being taken at fixed system size $L$.)

The rate function determines the probability of rare events where $c[\mathbf{x}]$ is non-typical, which means that the tagged particle is in a region where the density deviates significantly from its average, over a long time period.  Previous work has considered large deviations of the dynamical activity~\cite{Jack2015,thompson_dynamical_2015, DasLimmer2019,Dolezal_2019}, which are closely related to large deviations of $c[\mathbf{x}]$.  In particular, trajectories with high clustering have low dynamical activity.  

 However, we emphasise that the clustering considered here is a single-particle quantity, in contrast to previous work on the total activity of the system~\cite{Jack2015,thompson_dynamical_2015,Dolezal_2019}.

\subsubsection{Clustering in one dimension}

The clustering for the $1d$ model is measured over a length scale $a$ that is of the same order as the particle size $l_0$.  
In numerical work we take $(a/l_0) = 0.1$.  The clustering is defined as
\begin{equation}
	c_i(t)= 1-r^a_i
\label{equ:ci}
\end{equation}
where we note that $a$ is a label (not an exponent), and 
\begin{multline}
	r_i^a = \frac{1}{2a} \bigg[ \min(a,|x_i - x_{i+1}|-l_0 )\\ 
	+ \min(a,|x_i - x_{i-1}|-l_0 )  \bigg]
\end{multline}
is the fraction of free (unoccupied) space in a region of size $2a$, centred on the tagged particle.  (The parameter $a$ is the same as the maximal MC step, so that $r_i^a$ coincides with the probability that a trial move for particle $i$ is accepted, this allows a connection between clustering and dynamical activity, see~\cite{Dolezal_2019}.  The activity of particle $i$ considered in that work corresponds to $1-c_i$ in the current notation.)
Following~\cite{Dolezal_2019}, the equilibrium average of $c_i$ is

	\begin{equation}
	c_{\rm eq}(\rho) = \frac{\rho a-1+e^{-\rho a}}{\rho a} \; .
	\label{equ:ave-K}
	\end{equation}
At low density $\rho a \ll 1$,
 the clustering $c_{\rm eq} \approx \rho a/2\ll 1$.  For high density ($\rho\to\infty$) then $c_{\rm eq} \to 1$, which is the maximal possible value.

 \subsubsection{Clustering in two dimensions}
 \label{sec:cluster-2p}

To measure clustering in $2d$, we first define a measure of proximity between particles $i$ and $j$ as
\begin{equation}
Q_{i,j}=\begin{cases}
	\frac{1}{2} , &|  \vec{x}_i -\vec{x}_j|\leq \frac{l_0}{2}\\
	| \vec{x}_i-\vec{x}_j|/l_0, &\frac{l_0}{2} <| \vec{x}_i -\vec{x}_j|\leq l_0 \\
	0, &| \vec{x}_i -\vec{x}_j| > l_0 \; .
\end{cases}
\end{equation}
Considering the particles as discs with diameter $l_0$, this measures
 the extent to which two particles overlap each other, with a cutoff value of $\frac12$ when they get very close.

Then the clustering for particle $i$ is
\begin{equation}
c_i = \sum_{j(\neq i)} Q_{i,j}
\label{equ:ci-2d}
\end{equation}
(The sum runs over all particles $j$ except $j=i$.)

\subsection{Biased ensembles}
\label{sec:s-ens}

We investigate the rare (large-deviation) events associated with the distribution (\ref{equ:ldp}) using 
biased ensembles of trajectories, defined according to standard methods~\cite{Derrida2007,Lecomte2007,Chetrite2015}.
For the natural (unbiased) dynamics of the system, the average of an observable quantity $O$ is denoted by 
\beq
\label{equ:ave-eq}
\langle O \rangle_0 = \int \mathcal{D}\mathbf{x} \; O[\mathbf{x}] e^{-S_0[\mathbf{x}]}
\eeq
which is a path integral over all possible trajectories of the system, whose probabilities are given by the action $S_0[\mathbf{x}]$ which is defined as
\begin{equation}
	 S_0[\mathbf{x}]=\sum_i \int_{0}^{\tobs}\frac{(\dot x_i + \beta D_0 \nabla_i U)^2}{4D_0}  dt \; .
\end{equation}

Now consider an ensemble in which
the probability of trajectory $\textbf{x}$ is biased according to its clustering.  The probability density for this trajectory is 
\begin{equation}\label{sensemble}
P_{s}[\mathbf{x}]=\frac{1}{Z_s} P_0[\mathbf{x}]  e^{s\tobs c[\mathbf{x}]}  ,
\end{equation}	
where $P_0$ is the unbiased probability and the normalisation constant is
\begin{equation}
Z_s=  \left\langle {\rm e}^{s\tobs c[\mathbf{x}]} \right\rangle_0
\end{equation}	
Note that the biased ensemble (\ref{sensemble}) is defined so that $s>0$ corresponds to larger clustering.  Since clustering is anti-correlated with dynamical activity, this means that $s>0$ corresponds to reduced activity, similar to~\cite{Jack2015,thompson_dynamical_2015, DasLimmer2019,Dolezal_2019,Garrahan2009,Hedges2009}.

 By analogy with (\ref{equ:ave-eq}), averages in the biased ensemble are given by
\beq
\label{equ:ave-s}
\langle O \rangle_s = \frac{1}{Z_s}\int \mathcal{D}\mathbf{x} \; O[\mathbf{x}] e^{-S_0[\mathbf{x}] + 
s\tobs c[\mathbf{x}]}  \; .
\eeq

Investigating such biased ensembles provides insight into the rare fluctuations of the unbiased ensemble. One can invoke an analogy between these biased ensembles and canonical ensembles in statistical mechanics, see~\cite{Chetrite2013,Touchette2009} for a discussion. 
This motivates us to define the dynamical free energy,
\begin{equation}\label{equa::DFE}
\psi_L(s)=\lim\limits_{\tobs \rightarrow \infty} \frac{1}{\tobs} \ln Z_s \; .
\end{equation}
The rate function of (\ref{equ:ldp}) can be obtained from the free energy by Legendre transform $I_L(c) = \sup_s [ sc - \psi_L(s) ]$.
The dynamical analogue of the thermodynamic (average) energy density is
\begin{equation}\label{equa:cMeanDef}
	\langle {c}\rangle_s=\lim\limits_{\tobs \rightarrow \infty}\frac{1}{Z_s}\int \mathcal{D}\mathbf{x} \; c[\mathbf{x}] e^{-S_0[\mathbf{x}] + s\tobs c[\mathbf{x}]}
\end{equation} 
which satisfies $\langle {c}\rangle_s=\psi'_L(s)$.  The average is over trajectories of fixed length $\tobs$ so this quantity depends implicitly on $\tobs$, as well as the parameters of the model. 
The dynamical analogue of the specific heat capacity is the
asymptotic variance of the clustering 
\begin{equation}\label{equ:asympVar}
\chi_L(s)=\lim\limits_{\tobs \rightarrow \infty} \tobs\left(\langle {c^2}\rangle_s-\langle {c}\rangle_s^2\right)
\end{equation}
which satisfies $\chi_L(s)=\psi''_L(s)$.  For future reference it is useful to define the finite-time analogue of this quantity
\begin{equation}\label{equ:asympVar-finite-t}
\chi_{L}(s,\tobs)= \tobs\left(\langle {c^2}\rangle_s-\langle {c}\rangle_s^2\right) \; .
\end{equation}

\begin{figure*}	
	\centering
	\includegraphics[width=1\linewidth]{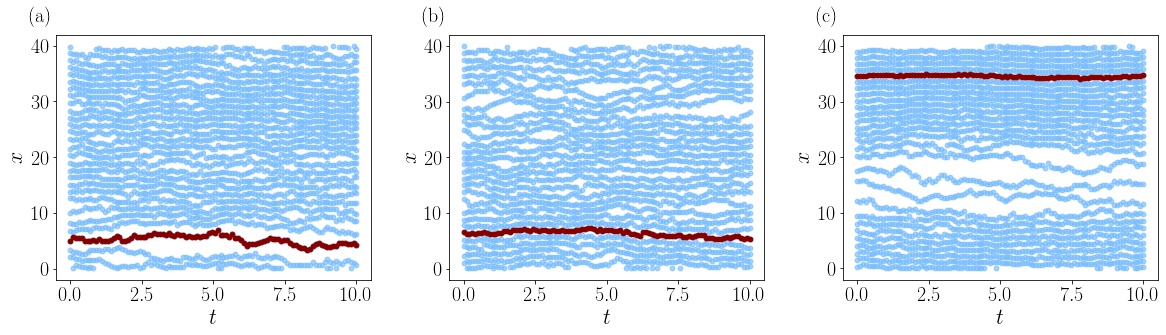}
	\caption{Typical trajectories from biased ensembles in one dimension.  (a) $\mu=-48$, biased to low clustering;
	(b) $\mu=0$, unbiased dynamics; (c) $\mu=48$, biased to high clustering.  The ensembles are biased by properties of a single tagged particle (shown in red), which elicits a system-wide response. Systems have $N=28$ particles at density $\rho_0=7/3$.  This same density is used for all numerical results in one-dimensional systems. The time is in units of $\tau_{\rm B}$ and $x$ is in units of $l_0$. }

	\label{fig::1DSnapshot}
\end{figure*}

\subsection{Biased ensembles in the thermodynamic limit: long-ranged correlations and hydrodynamics}

The LDP (\ref{equ:ldp}) describes the probabilities of rare events in the limit $\tobs\to\infty$.  The rate function for this LDP depends on the system size $L$.  Our focus will be on the nature of these rare events for large systems, that is $L\to\infty$ with a fixed (overall) density $\rho_0$.  

Recall that the clustering $c_i$ in (\ref{equ:cbar}) only depends on a single particle and its local environment.  In this case, one might expect that the fluctuations of this particle are independent of the system size, in which case there would be a non-trivial limiting rate function $I_\infty(c)=\lim_{L\to\infty} I_L(c)$.
  
This paper shows that this (very natural) expectation is wrong for the systems under consideration, and we argue that this feature is rather general.
 Instead, the large-deviation events associated with (\ref{equ:ldp}) depend significantly on the system size and involve correlations of the tagged particle with all the other particles in the system.  In terms of the rate function, we find for $d=1$ and $L\to\infty$ 
\begin{equation}
I_L(c) \approx \frac{1}{L} \mathcal{I}(c)
\end{equation} 
with ${\cal I}(c)=O(1)$, so large-deviation events are less rare in larger systems.  (The approximate equality is accurate when $L$ is large.)  The corresponding result for the dynamical free energy is obtained by Legendre transform,
\begin{equation}
\psi_L(s) \approx \frac{1}{L} \Psi(sL) \; ,
\label{equ:psi-L-1d}
\end{equation}
valid for $sL=O(1)$ as $L\to\infty$.

Physically, these scaling forms appear because large-wavelength density fluctuations in the system are associated with very long (hydrodynamic) time scales, which couple strongly to time-averaged quantities like (\ref{equ:cbar}).  Since the hydrodynamic modes involve long-ranged correlations of the density, this means that biased ensembles such as (\ref{sensemble}) are characterised by system-wide responses to the bias, even though  $c[\vec{x}]$ is a single-particle quantity.

To see how hydrodynamic time scales affect the large deviations, note from (\ref{equ:psi-L-1d}) that $\chi_L(0)=\psi''_L(0)=O(L)$.  Hence by (\ref{equ:asympVar}), the asymptotic variance of the clustering diverges in large systems.  Using (\ref{equ:cbar}) with (\ref{equ:asympVar}) one sees that
\begin{equation}\label{equ::OtherVarDef}
	\chi_L(s)= \int_{-\infty}^{\infty} \left[ \langle c_i(t)c_i(0) \rangle_s - \langle c\rangle_s^2 \right] \mathrm{d}t \; .
\end{equation}
The integrand is a correlation function that is always less than unity, so the divergence of $\chi_L(0)$ as $L\to\infty$ must be due to slowly decaying correlations, which originate in hydrodynamic modes, as we shall see below.

\section{Biasing one particle in one dimension}\label{sec::Results1D}

We consider the system in $d=1$.
On large scales, the hydrodynamic behavior is that of a diffusing density field.  There is a corresponding hydrodynamic time scale 
\begin{equation}
\tau_{L} = \frac{ L_r^2 }{ 2D_0 } \; ,
\end{equation}
 which diverges with the system size.  
 When considering biased ensembles of trajectories, it is useful to define a dimensionless observation time
\beq\label{eqn:gammaobs}
\gammaobs  = \frac{\tobs}{\tau_L} \; .
\eeq
In numerical work, we compare different system sizes at fixed $\gammaobs$ (as well as fixed density $\rho_0$).   This requires very large observation times $\tobs$ when considering large systems, but is essential for a meaningful finite-size scaling analysis.  {For the large-deviation limits of (\ref{equ:ldp},\ref{equa::DFE}), the parameter $\gammaobs$ should also be large.}
Anticipating the scaling form (\ref{equ:psi-L-1d}), it is useful to define a rescaled (and dimensionless) biasing field
\beq
\mu  = \frac{ s\LL a}{ D_0 } \; .
\label{equ:def-mu}
\eeq

We present numerical results obtained by transition path sampling (TPS), 
as implemented in~\cite{Dolezal_2019}.  This enables sampling of representative trajectories from (\ref{sensemble}), for ensembles with prescribed $L,\tobs$.
Fig. \ref{fig::1DSnapshot} shows representative trajectories for different values of $\mu$, corresponding to clustering that is larger or smaller than average.  It is clear that biasing the tagged particle leads to a response that spans the whole system -- the particle is either embedded in a macroscopic cluster of particles, or in a macroscopic void with very few particles.

\subsection{Hydrodynamic theory}
\label{sec:hydro-1}

To understand the behavior shown in Fig.~\ref{fig::1DSnapshot},
we outline a fluctuating hydrodynamic theory for this system~\cite{Bertini2005,Dolezal_2019,Jack2015}.  This establishes the scaling of (\ref{equ:psi-L-1d}), and clarifies its relationship with long-ranged density fluctuations.

The hydrodynamic theory is expressed in terms of a coarse-grained density field which is defined by averaging over a mesoscopic spatial region of size $\Omega$:
\beq
\rho(x,t) = \frac{1}{\Omega} \int_{x-\Omega/2}^{x+\Omega/2} \sum_i \delta(y-\tilde{x}_i(t)) {\rm d}y
\label{equ:def-rho-field}
\eeq
Note we use co-ordinates in the reduced representation, so $x\in [0,\LL)$.    
The dynamics of $\rho$ obeys a continuity equation
 \begin{equation}\label{equ::langevin}
\partialD{\rho}{t}=-\nabla\cdot j \; .
\end{equation}
Given the diffusive dynamics of the system, it it is consistent to assume that
\begin{equation}
  \label{equ::langevin2}
 j=-D(\rho)\nabla\rho+\sqrt{2\sigma(\rho)} \,\eta
 \end{equation}
where $D(\rho)$ and $\sigma(\rho)$ are a diffusivity and a mobility, and $ \newcommand{\e}{{\rm e}} \eta$ is a space-time white noise with mean zero and $ \newcommand{\e}{{\rm e}} \langle \eta(x,t) \eta(x',t')\rangle = \delta(x-x') \delta(t-t')$. These equations can be derived by considering the empirical density and current and taking the hydrodynamic limit,~\cite{dean_langevin_1996}.

In general, we note this hydrodynamic framework (which comes from MFT) is a robust and general theory for diffusive systems. The functions $D,\sigma$ depend on microscopic details of the system (for example, the interaction potential between particles), but in many cases, the dependence of these functions on $\rho$ cannot be derived from first principles.  However, one may still derive and exploit generic predictions for hydrodynamic behaviour, so that many different aspects of the system behaviour can all be expressed in terms of these two functions.

When considering the density field, the particles are indistinguishable.  As discussed in~\cite{Dolezal_2019}, this means (for the specific system considered here) that trajectories in this reduced representation are in one-to-one correspondence with an ideal gas of diffusing particles, and hence
\beq
D(\rho)=D_0, \qquad \sigma(\rho) = \rho D_0 \;,
\label{equ:D-sig-bhpm}
\eeq 

Since this theory is valid on hydrodynamic time scales, it is also convenient to define rescaled coordinates
\beq
\hat{x}=\frac{x}{\LL}, \qquad \hat{t}=\frac{t}{\tau_{L}} ,
\eeq
and a rescaled current $\hat\jmath = {j\tau_{L}}/{\LL} $.  (Note however that the density is \emph{not} rescaled.)

\begin{figure}
	\centering
	\includegraphics[width=1\linewidth]{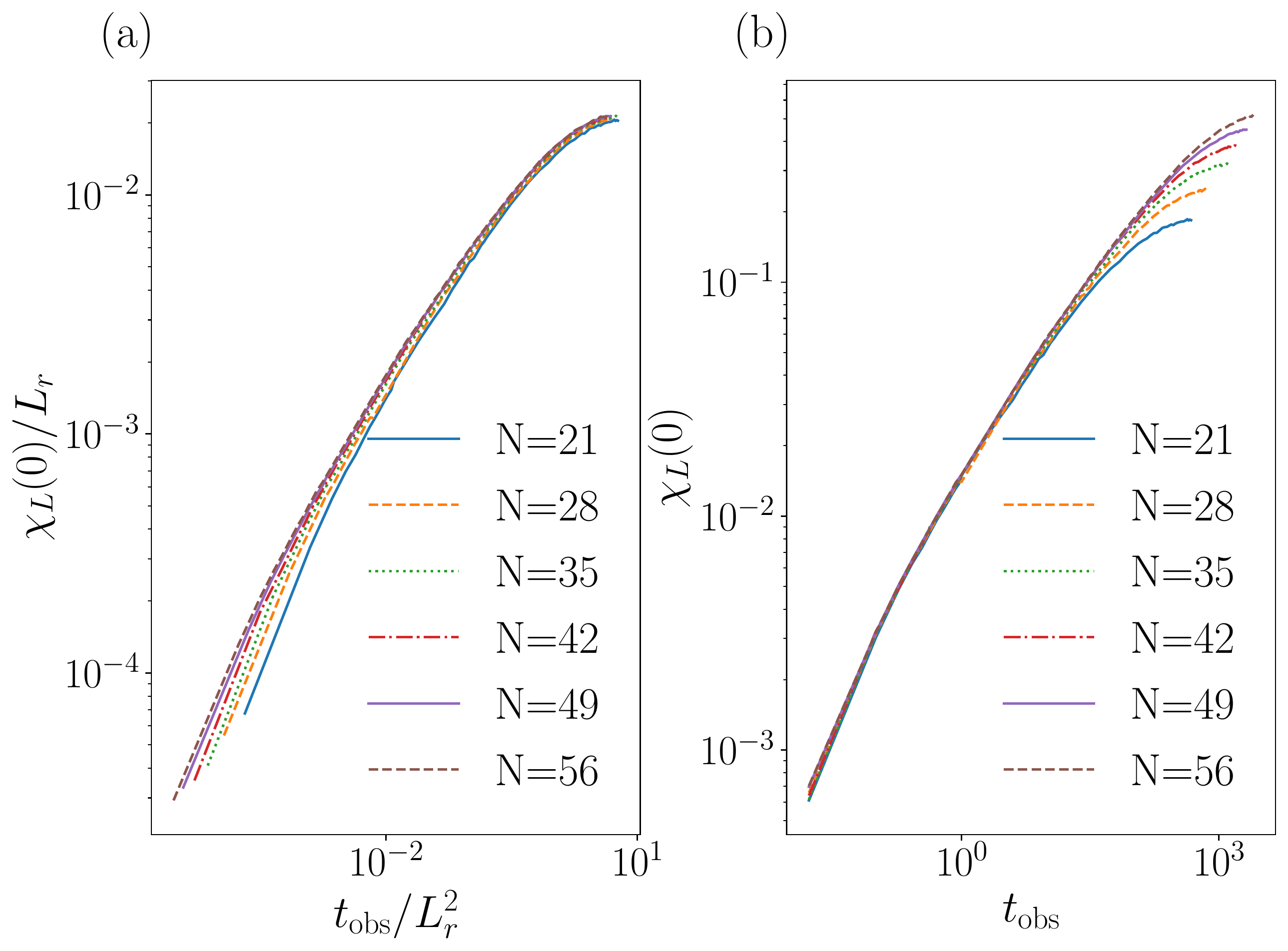}
	\caption{Scaled variances of $c[\vec{x}]$ in unbiased dynamics, as defined in (\ref{equ:asympVar-finite-t}), for increasing system sizes.  Panels (a) and (b) illustrate the slow and fast scaling regimes of (\ref{equ:chi-f}) respectively. The average density is $\rho_0=7/3$. 
	}
	\label{fig:1DVariance}
\end{figure}

\subsection{Hydrodynamic scaling for typical fluctuations}

To connect this theory with fluctuations of $c[\textbf{x}]$, it is useful to decompose $c_i(t)$ into slow (hydrodynamic) and fast contributions as
\beq
\label{eqn:cslow+fast}
c_i(t) = c_i^{\rm slow}(t) + c_i^{\rm fast}(t)
\eeq
where $c_i^{\rm slow}=c_{\rm eq}(\rho(x_i))$ is the average clustering in a system whose density is $\rho(x_i)$ [recall (\ref{equ:ave-K})], while $c^{\rm fast}$ is the remainder.   Writing $\rho(x_i)=\rho_0 + \delta\rho(x_i)$, the typical size of the fluctuations is $\delta\rho = O(L^{-1/2})$.  (We assume the mesoscopic length $\Omega =O(L)$ as we take the hydrodynamic limit.)  Then $c^{\rm slow}_i \approx c_{\rm eq}(\rho_0) + \delta\rho(x_i) c_{\rm eq}'(\rho_0)$.  For a simple analysis, assume that $c^{\rm fast}$ and $c^{\rm slow}$ are statistically independent; then (\ref{equ:asympVar-finite-t}) becomes
\begin{multline}
\chi_{L}(0,\tobs) = 
\frac{c'_{\rm eq}(\rho_0)^2}{\tobs}  \operatorname{Var}\left[ \int_0^{\tobs} \delta\rho(x_i(t)) {\rm d}t \right]
\\ + 
\frac{1}{\tobs} \operatorname{Var}\left[ \int_0^{\tobs} c^{\rm fast}(t) {\rm d}t \right] \; .
\label{equ:chi-fast-slow}
\end{multline}
The object in the second line is a fast contribution which is independent of $L$, it increases as a function of $\tobs$ and saturates to a limiting value $\chi_{\rm fast}$.  Such a contribution is present in all systems and intuitively would be the only one present in a simple system without slow hydrodynamic modes.  By contrast, the first line is specific to systems with hydrodynamic modes: the typical size of the density fluctuation is $O(L^{-1/2})$ so this contribution is $O(1/L)$ for short and moderate $\tobs$, hence negligible with respect to the fast term.  However, the slow (diffusive) relaxation of $\rho$ means that the variance in the first line is a scaling function of $\tobs/\tau_L$, so for $\tobs\to\infty$ it scales as $\tau_L \times O(1/L)=O(L)$. This separation of timescales is related to the onset of local equilibrium in fluids  first remarked upon by Bogoliubov~\cite{Bogoliubov, ORTIZDEZARATE2006113}. On time scales where the average distance diffused by a particle is smaller than the interparticle separation, local equilibrium is not established yet and the variance follows $c^{\rm fast}$. When the average distance diffused becomes larger than the interparticle separation, local equilibrium is established. The particle then is affected by the entire system and the clustering scales as $c^{\rm slow}$. Hence indeed $\chi_L(0)=O(L)$, consistent with (\ref{equa::DFE},\ref{equ:asympVar}).

To summarise, the fast term dominates for $\tobs=O(1)$ while the slow term dominates at large times:
\beq
\label{equ:chi-f}
\chi_L(0,\tobs) = \begin{cases}
f_{\rm fast}(\tobs), & \tobs = O(1), \\
\LL f_{\rm slow}(\tobs/\tau_L), & \tobs = O(\LL^2) \; .
\end{cases}
\eeq
where $f_{\rm fast},f_{\rm slow}$ are scaling functions. 
Fig.~\ref{fig:1DVariance} illustrates these two scaling regimes.  

The conclusion of this section is that understanding \emph{typical} fluctuations of $c[\textbf{x}]$ already requires an analysis of hydrodynamic modes.  We have shown that the asymptotic variance of this single-particle quantity diverges with system size.  The consequences of hydrodynamic modes for large deviations will be discussed next.

\subsection{Hydrodynamic theory of large deviations}

\begin{figure}
	\centering
	\includegraphics[width=1\linewidth]{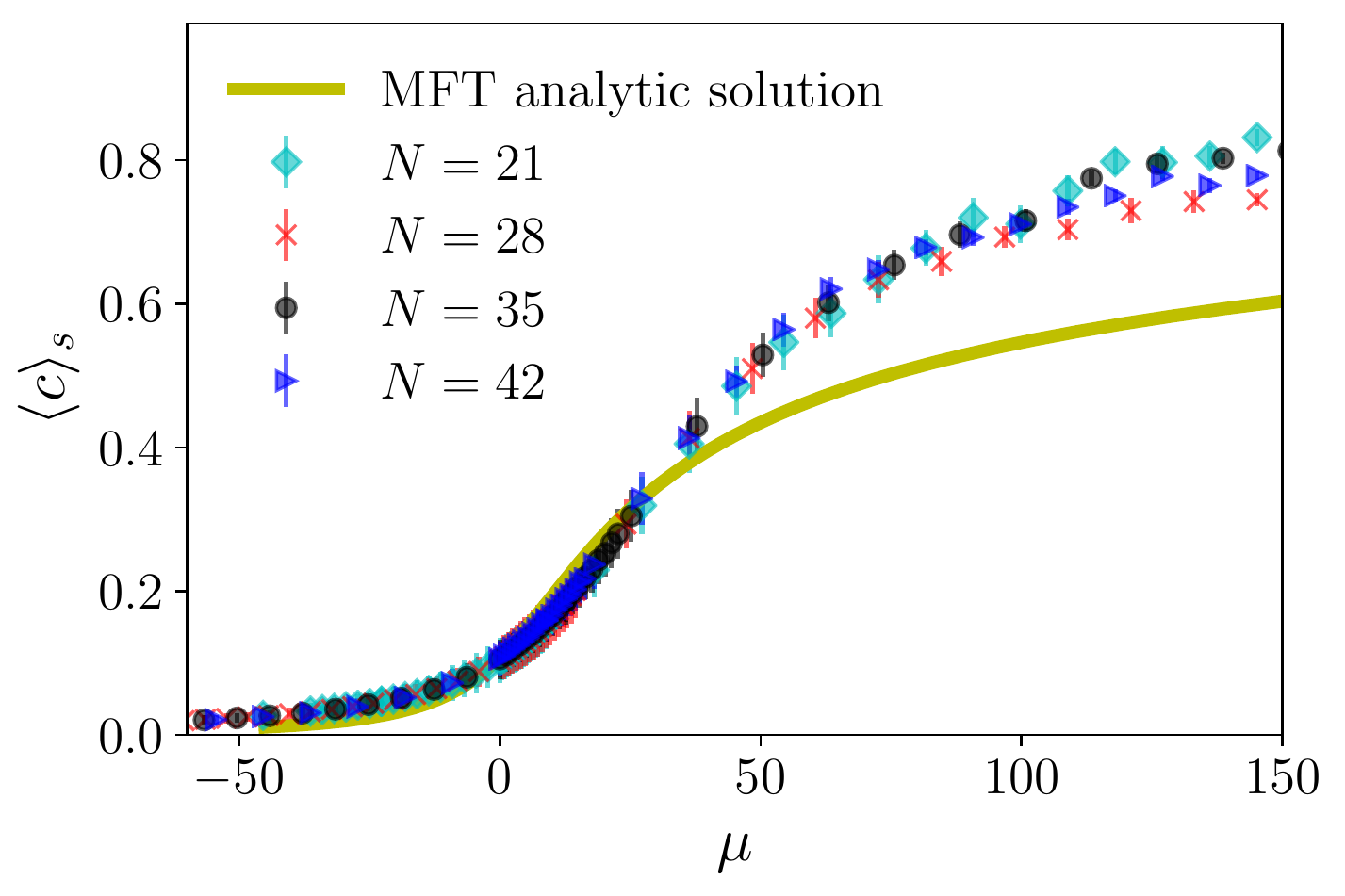}
	\caption{Mean clustering $\langle c\rangle_s=\psi_L'(s)$, plotted as a function of the scaling variable $\mu$ for different system sizes.  (Here and throughout, symbols show numerical results for the biased ensemble described by (\ref{equ:ave-s}).)  The scaling collapse illustrates the form \refequa{equ:psi-L-1d}. The analytic solution is based on the minimisation problem (\ref{equ:psi-var-1d}), which is solved by minimising (\ref{equ:analySol},\ref{equ:analySol-v2}) over $\chi$ and $\chi_1$ respectively to obtain the density profile $\rho$.  Using this calculation the corresponding activity is obtained from (\ref{equ:ave-K}) with $\rho=\rho_p$. We take $\gammaobs  =0.07$ and $\rho_0=7/3$. }
	\label{fig::1DMeans}
\end{figure}

We formulate a path integral for the density and current.  Since the system also includes a tagged particle, this has to be considered separately in the path integral.  However, since particles cannot pass each other, the typical displacement of the tagged particle is subdiffusive~\cite{Dhar_2014,Krapivsky2014}, and it is consistent to treat that particle as stationary on the hydrodynamic scale, with a fixed position $X_{p}$.  We also define $\rho_p = \rho(X_{p})$ as the density in the vicinity of the tagged particle, and we estimate the time-averaged clustering as
 \beq
c[\vec{x}] = \frac{1}{\gammaobs}\int_0^{\gammaobs} c_{\mathrm{eq}}(\rho_p) d\hat{t}
\label{equ:c-hydro}
 \eeq
where $c_{\mathrm{eq}}(\rho)$ was defined in (\ref{equ:ave-K}) as the average clustering for a particle in equilibrium at density $\rho$.  The estimate (\ref{equ:c-hydro}) is accurate in the hydrodynamic limit because the particle explores its (locally-equilibrated) environment on a time scale that is fast compared with $\tau_L$.  It amounts to replacing $c_i$ by $c_i^{\rm slow}$, which is valid because the slow modes are the dominant source of fluctuations when $\tobs$ is large, recall (\ref{equ:chi-fast-slow}).

Then, the path integral on the hydrodynamic scale, analogous to (\ref{equ:ave-s}), is
\beq
\langle O \rangle_s = \frac{1}{{\cal Z}_s} \int \mathcal{D}(\rho,j) \; O[\rho,j] e^{-S[\rho,j, x_c] } \delta\left(\frac{d\rho}{d\hat t}+\nabla\cdot \hat \jmath \right)
\label{equ:ave-s-hydro}
\eeq
where the delta function
 restricts the integral to paths that obey the continuity equation (\ref{equ::langevin}), and ${\cal Z}_s=\int \mathcal{D}(\rho,j) \;e^{- S[\rho,j, x_p] }\delta(\frac{d\rho}{d\hat t}+\nabla \cdot \hat \jmath)$ is a normalisation constant.  Also 
 \begin{multline}
 \label{equ:action-rho-j}
S[\rho,j, x_p] = \LL  \int_0^{\gammaobs}  \int_0^{1} D_0\frac{|\hat \jmath +\hat \nabla\rho/2|^2}{2\sigma(\rho)} \, \mathrm{d}\hat{x}\mathrm{d}\hat{t}
 \\-  \frac{\mu\LL}{2a}  \int_0^{\gammaobs}  c_{\mathrm{eq}}(\rho_p) \mathrm{d}\hat{t}
 \end{multline}
 where boundary terms have been neglected.
The first line of this expression is familiar from Macroscopic Fluctuation Theory~\cite{Bertini2005}, the second uses (\ref{equ:c-hydro}).

Note that (\ref{equ:ave-s}) and (\ref{equ:ave-s-hydro}) are different ways of defining biased ensembles of trajectories: the first is used for particle models and the second for hydrodynamic theories (MFT).  To the extent that the hydrodynamic theories represent accurately the underlying particle models, we expect the two definitions to result in the same behaviour.  The numerical simulations in this work are all obtained for particle models, using (\ref{equ:ave-s}).  The MFT results consistently use  (\ref{equ:ave-s-hydro}).  Several figures (including for example Fig.~\ref{fig::1DMeans}) test the extent to which the two descriptions agree with each other.

\begin{figure}
	\centering
	\includegraphics[width=1\linewidth]{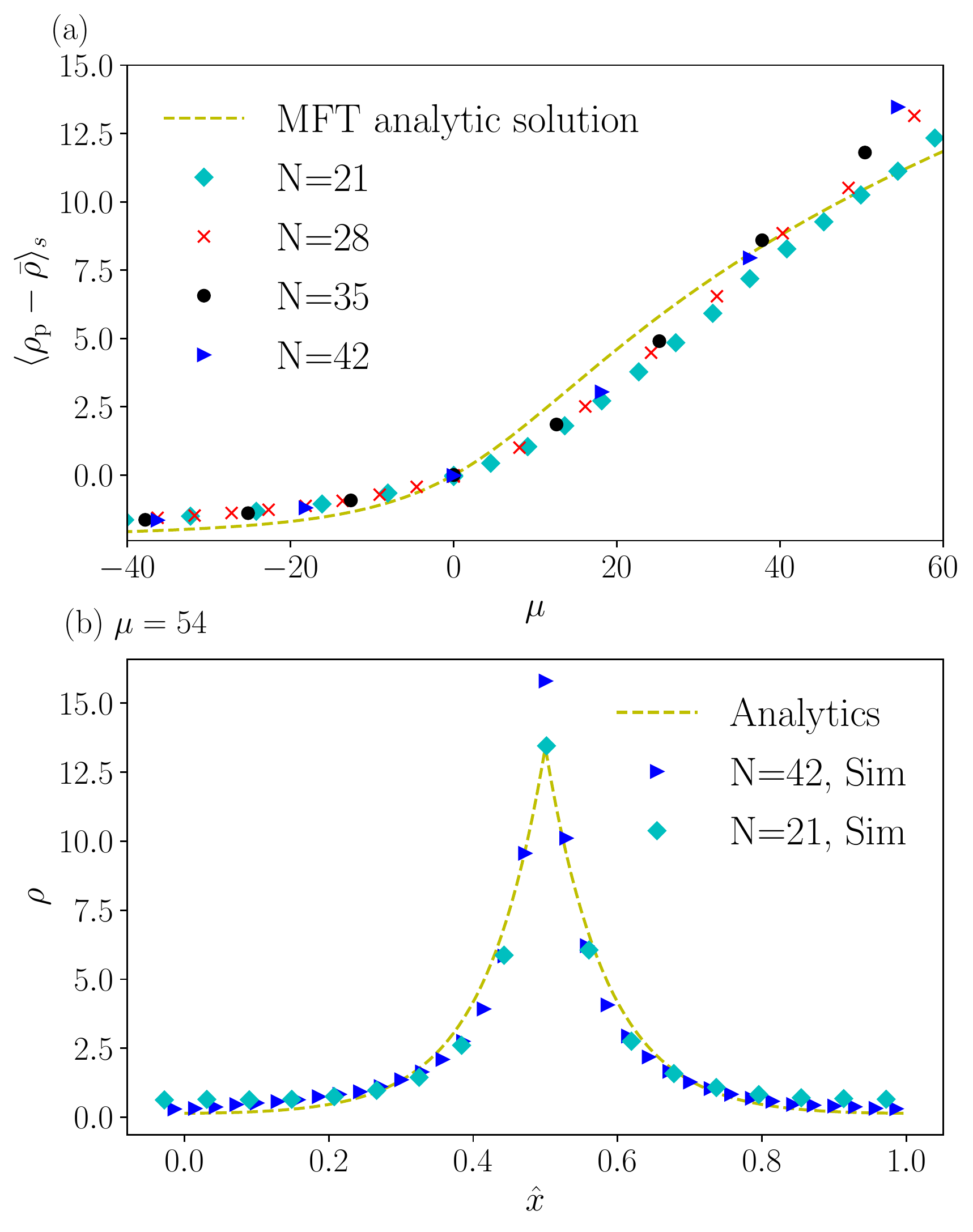}
	\caption{(a) Average density $\rho_p$ as a function of the bias $\mu$ from numerical simulations, compared with the analytic solution of the variational problem (\ref{equ:psi-var-1d}), see (\ref{equ:analySol},\ref{equ:analySol-v2}).
	(b) The corresponding density profiles: 
	numerical results have system size $N=42$ and $\mu=54$. In this figure we take $X_p=0.5$, the analysis in the main text uses $X_p=0$, which can be recovered by a simple translation in space. We take $\gammaobs  =0.07$ and $\rho_0=7/3$. 
	}
	\label{fig:maxdensitysimulations}
\end{figure}

At hydrodynamic level, the dynamic free energy is $\psi_L(s)=\lim_{\tobs\to\infty} (1/\tobs) \ln {\cal Z}_s$, analogous to Eqn. \refequa{equa::DFE}. For large $L$, this may be obtained by a saddle-point method. For $s=0$ the action is minimised by a constant density $\rho=\rho_0$ with $j=0$. The bias $s$ does not break time-reversal symmetry so one has generally that  $j=0$ and that the optimal trajectory is constant in time, but the optimal $\rho$ may depend on $x$. Hence by minimising the action one has for large $L$ that
 \begin{equation}
\psi_L(s)= \frac{-D_0}{\LL}\min_{\rho(\hat{x})}\bigg[\int_{0}^{1}\frac{|\hat\nabla\rho|^2}{4\rho}\mathrm{d}\hat x - \frac{\mu c_{\mathrm{eq}}(\rho_p)}{a}  \bigg]
\label{equ:psi-var-1d}
\end{equation}
where we used (\ref{equ:D-sig-bhpm}) to substitute for $\sigma(\rho)$.
The value of $X_p$ is irrelevant for the minimization.

This minimization problem will be considered next, but it is already clear from this expression that $\psi_L$ obeys the scaling form (\ref{equ:psi-L-1d}).

Numerical evidence for this scaling form is shown in Fig.~\ref{fig::1DMeans} which shows $\langle c \rangle_s = \psi'(s)$, which is indeed of the form $\langle c \rangle_s = f_c(\mu)$, where  $f_c$ is a scaling function and the scaling variable $\mu$ was defined in (\ref{equ:def-mu}).

We now solve the variational problem (\ref{equ:psi-var-1d}). The position $\hat x\in [0,1)$ is measured relative to the tagged particle, so $X_p=0$.  Considering the functional derivative with respect to $\rho(\hat{x})$ one obtains (for $\hat{x}\neq0$):
\begin{equation}
	-\frac{\hat\nabla^2\rho(\hat x)}{2\rho(\hat x)}+\frac{(\hat\nabla\rho(\hat x))^2}{4\rho(\hat x)^2}=-\chi^2,
	\label{equ:HJ-rho}
\end{equation}
where  $\chi^2$ is a lagrange multiplier that ensures conservation of the number of particles, $\int_0^1 \rho(\hat{x})d\hat{x} = \rho_0$. For $\mu>0$, the general solution of (\ref{equ:HJ-rho}) is 
\begin{equation}
\rho(\hat x)=A \left[ \cosh(2\chi(B+\hat x))+1\right]
\label{equ:cosh}
\end{equation}
with integration constants $A,B$.

To gain insight into the minimization problem (\ref{equ:psi-var-1d}), it can be useful~\cite{lecomte_inactive_2012} to draw an analogy between the spatial profile $\rho(x)$ and a trajectory of a particle in classical dynamics.  The analogy identifies $x$ and $\rho(x)$ with $t$ and $q(t)$, where $t$ is a time variable and $q$ a position co-ordinate.  So $\nabla\rho$ is identified with the velocity $\dot q(t)$.  Within this mapping then (\ref{equ:psi-var-1d}) becomes the classical action for a dynamical system, expressed as the time integral of a Lagrangian ${\cal L}(q,\dot q,t)$ where (assuming as before that the probe particle is localised at the origin) the explicit time-dependence of the Lagrangian appears in a term $-\frac{\mu c_{\rm eq}(q)}{a}\delta(t)$.

 Incorporating as before the normalization constraint on $\rho$ by the Lagrange multiplier $\chi^2$, the corresponding Hamiltonian can be constructed as ${\cal H}(q,\pi,t)=q\pi^2+\chi^2q+\frac{\mu c_{\rm eq}(q)}{a}\delta(t)$ where $\pi = \partialD{\cal L}{\dot q} =\dot q/(2q)$ is the momentum variable conjugate to $q$.  Since the Hamiltonian only depends on $t$ via the delta function at $t=0$, the energy is conserved along the trajectory (except at $t=0$).  Such an approach allows the second-order differential equation (\ref{equ:HJ-rho}) to be converted to first-order form, with the value of the energy entering as one of the integration constants.  The value of this energy can be obtained by using the fact that the action is to be minimized over periodic trajectories (with a period of unity), because of the periodic boundaries of the original spatial model.
 
 In the present context, this analogy with dynamical systems provides one possible way to derive (and interpret) the general solution (\ref{equ:cosh}), as well as the corresponding result (\ref{equ:cos}) below, which is relevant when $\chi^2<0$.

 
 With the solution (\ref{equ:cosh}) in hand, we next observe that the minimiser of (\ref{equ:psi-var-1d}) is symmetric about $\hat{x}=\frac12$ and has $\rho(0)=\rho(1)$ from which we find $B=-1/2$.  For $\mu>0$ then $\rho(0)>\rho_0$ which corresponds to $\chi^2>0$.  Enforcing the constraint on the mean density we find
\begin{equation}
A=\frac{\rho_0\chi}{\chi+\sinh\chi}
\label{equ:ABC-rho}
\end{equation}
The object to be minimised in (\ref{equ:psi-var-1d}) becomes 
\begin{multline}\label{equ:analySol}
	\frac{S}{\LL\tobs D_0} =
	\chi A(\sinh\chi-\chi) \\ + \mu/a\frac{A a \cosh\chi -1+\exp(-A a \cosh\chi)}{A a \cosh\chi}
\end{multline}
 This expression can be minimised numerically over $\chi$ subject to (\ref{equ:ABC-rho}) and hence a density profile for $\rho$ is found.  (In the analogy with the classical Lagrangian, this minimization accounts for the effects of the delta-forcing term in  ${\cal H}$ at $t=0$.)
 
 For $\mu<0$ it is necessary to take $\chi^2<0$ so we define $\chi_1=-{\rm i}\chi$. \refequa{equ:HJ-rho} then becomes
 \begin{equation}
 	\rho(\hat x)=A \left[ \cos(2\chi_1(B+\hat x))+1 \right]
	\label{equ:cos}
 \end{equation} The equation for A becomes
 \begin{equation}
 	A=\frac{\rho_0\chi_1}{\chi_1 + \sin\chi_1}
 	\label{equ:ABC-rho-neg}
 \end{equation}

The expression for $S$ with negative bias becomes
\begin{multline}\label{equ:analySol-v2}
	\frac{S}{\LL\tobs D_0} =\chi_1 A(\chi_1-\sin\chi_1) \\ + \mu/a\frac{A a \cos\chi_1 -1+\exp(-A a \cos\chi_1)}{A a \cos\chi_1}
\end{multline}
A similar situation with cosine and cosh profiles was found in~\cite{keta2020collective} for the distribution of two particles that are biased to be more or less frequently in contact with each other.

\begin{figure*}
	\centering
	\includegraphics[width=1\linewidth]{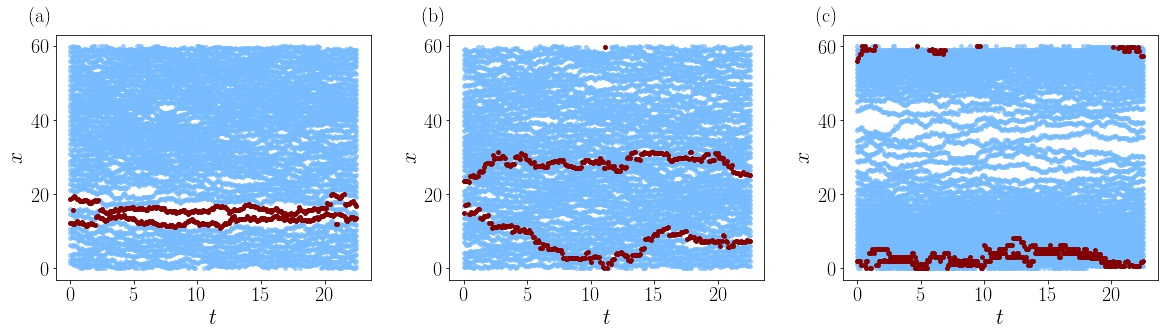}
	\caption{Typical trajectories for different levels of bias with two tagged particles in a system with $N=42$ at (a) $\mu=-10.6$, (b) $\mu=0$ and (c) $\mu=10.6$.  (a) With large negative bias the tags on the particles are localized in a region of low density.  (b) With no bias the tags diffuse freely around the system.  (c) With  large positive bias the tags diffuse in a very dense region. The density of the system is $\rho_0=7/3$ and we take $\gammaobs=0.07$. The time is in units of $\tau_{\rm B}$ and $x$ is in units of $l_0$. }
	\label{fig:Snapshot1D2Part}
\end{figure*}

\subsection{Comparison of analytic and numeric results}

We now compare our analytical predictions with results from simulations.  We emphasise that our theoretical approach requires that $L$ is large (to justify the hydrodynamic approach) and also that $\gammaobs\to\infty$, to ensure that trajectories are longer than the relevant relaxation times (including $\tau_L$), so that the system is in the large-deviation regime.  For numerical simulations, the results are obtained by TPS, which limits the accessible values of $L$ and $\gammaobs$.  We perform finite-size scaling in $L$ with fixed density $\rho_0$ and fixed $\gammaobs$: this already necessitates trajectories of length $\tobs \sim L^2$, due to (\ref{eqn:gammaobs}).  In these one-dimensional systems, we find that taking $N=\rho_0 L$ in the range $20-40$ is sufficient to observe clear signatures of hydrodynamic scaling.  
In all numerical results, we emphasise that density $\rho_0$ is fixed, so that $L$ and $N$ are both increasing together.  Hence finite-size scaling arguments can be equivalently expressed in terms of increasing $L$ or increasing $N$.
For the trajectory length, numerical results are obtained at $\gammaobs=0.07$: while this is not numerically large, previous work~\cite{Dolezal_2019} found this value sufficient for semi-quantitative agreement with predictions of the hydrodynamic large deviation theory, at manageable computational cost.  Still, one should bear in mind that neither $L$ nor $\gammaobs$ are large enough to expect fully quantitative agreement between theory and simulation.
 
Results for the dependence of $\langle c\rangle_s$ on $s$ and $L$ were shown already in Fig.~\ref{fig::1DMeans} above, showing the data collapse as a function of the scaling variable $\mu$ from (\ref{equ:def-mu}).  Comparing now with the hydrodynamic theory,
the analytical results agree with the data for small bias: there are small differences which can be attributed primarily to the finite values of $L$ and $\gammaobs$ used in the numerical simulations.

For larger positive bias, the agreement worsens: this may be partly attributable to deviations from the hydrodynamic theory when density gradients become too large, see for example~\cite{Dolezal_2019}. This effect is in addition to the finite size issues mentioned above which are accentuated at larger $\mu$.  We also checked that the analytical results (\ref{equ:cosh},\ref{equ:cos}) agree with direct numerical optimisation of  (\ref{equ:psi-L-1d}), via spatial discretisation of $\rho(x)$.

Analytical and numerical results for the density in the vicinity of the tagged (probe) particle are shown in Fig. \ref{fig:maxdensitysimulations}, including its dependence on $\mu$.  The density profile is also shown, as a function of the distance from the probe. There is semi quantitative agreement between the numerical and analytical results. 
As before, we attribute this to corrections from finite $L$ and $\gammaobs$.
An additional caveat is that the theory assumes that the probe particle is stationary, and that the presence of the probe does not itself affect the local density.  Both of these assumptions should be valid in large enough systems, but there will be deviations at finite $L$: these are examples of effects that contribute to deviations between theoretical and numerical results.

\subsection{Discussion -- long-ranged response to localised bias}

There are two important results of the analysis so far.  Firstly, Eq.~(\ref{equ:psi-var-1d}) and Fig.~\ref{fig::1DMeans} show that the response to the bias $s$ depends on the scaling parameter $\mu \propto sL$.  This means that $\psi_L(s)$ does not converge to a smooth (analytic) function as $L\to\infty$.   The second important result is Fig.~\ref{fig:maxdensitysimulations}(b) which shows the physical reason for the scaling behaviour: the bias acts on one particle but it generates a response in the particle density that covers the whole system, even as $L\to\infty$.

In the analogy between thermodynamic ensembles and biased trajectory ensembles~\cite{Touchette2009, Lecomte2007, Garrahan2009}, $\psi_L(s)$ corresponds to the free energy, which would be expected to have an analytic limit as $L\to\infty$ (unless the system is exactly at a phase transition).  Similarly, in thermodynamic systems, applying a localized bias would be expected to yield a localized response, except in unusual circumstances involving phase transitions or phase coexistence.  

Hence, our results show that properties of trajectory ensembles cannot always deduced from the analogy with thermodynamics.  
In the dynamical context, hydrodynamic modes can mediate long-ranged responses to localized bias, and to non-analytic free energies.
For systems with such modes, we conclude that some care is required when making analogies between thermodynamic ensembles and biased trajectory ensembles.  
We are not aware of analogs of these hydrodynamic effects in the thermodynamic context.

It is also useful to compare the results so far with trajectory ensembles that are biased by the total activity~\cite{appert-rolland_universal_2008,Jack2015,Dolezal_2019,YongJooDynamicalSymmetryBreaking}, instead of biasing a single particle.   In that case the free energy $\psi$ is a scaling function of $sL^2$ (instead of $sL$) and the scaling function has an additional singularity which corresponds to a spontaneous symmetry breaking, where the system becomes inhomogeneous.  In the present context, translational symmetry is explicitly broken by the choice of probe particle, and the system becomes inhomogeneous as soon as $\mu\neq 0$.   From our results, the scaling function $\Psi$ in (\ref{equ:psi-L-1d}) does not appear to have any singularity.
\section{Biasing two particles in one dimension}\label{sec::Results1D2p}

Given that long-ranged correlations appear on biasing a single tagged particle, one may expect interesting collective effects between multiple tagged particles, see also~\cite{GarrahanPretransition}.  We illustrate this by considering two tags.  
Since particles cannot pass each other in this $1d$ system, it is convenient to consider tags that can be transferred between different particles.  Specifically, if a tagged particle is separated by less than $a$ from an untagged particle, the tag is transferred between them with rate $D_0/l_0^2$.  (This rate is chosen so that the diffusive motion of the tag in a dense region of the system is comparable to the diffusion constant of a single particle in a dilute region.) 

Let the indices of the tagged particles be $p_1(t)$ and $p_2(t)$, so $c_{p_1(t)}(t)$ is the analogue of $c_i(t)$ considered in Sec.~\ref{sec::Results1D}.
Then the analogue of (\ref{equ:cbar}) is 
\begin{equation}
	c[\vec{x}]=\frac{1}{\tobs}\int_{0}^{\tobs} \left[ c_{p_1(t)}(t) +c_{p_2(t)}(t) \right] \mathrm{d}t \; .
\end{equation}

Representative trajectories in ensembles biased by this $c[\vec{x}]$ are shown in Fig. \ref{fig:Snapshot1D2Part}, which may be compared with Fig.~\ref{fig::1DSnapshot}.
For the unbiased (equilibrium) ensemble both the tags and the particles diffuse freely. 
On biasing the tagged particles to increased clustering, the tags tend to localise in a single dense region, since this is an efficient way for both tagged particles to have large $c_p$.   Similarly, on biasing the system to reduced clustering, the two tags tend to localise in a single region of low density, this has some similarities with~\cite{GarrahanPretransition}.

\subsection{MFT analysis}

It is striking that the two tags evolve almost independently in the unbiased system but they become strongly correlated when the bias is introduced.
 To analyse this, we generalise the action (\ref{equ:action-rho-j}) to include the positions of the tagged particles: the resulting action is
 $S = \int_0^{\gammaobs} {\cal L} {\rm d\hat t}$ with
\begin{multline}
	{\cal L}[\rho,\hat\jmath,\hat X_1,\hat X_2] =
	\LL \int_0^{1} 
	 \frac{(\hat{\jmath}+\hat\nabla\rho/2)^2}{2\rho}  \, \mathrm{d}\hat{x} \, 
	 \\
	 - \frac{\LL\mu}{2a}c(\rho_1,\rho_2)
	 + \frac{D_0}{2D_t} \left( \dot{\hat X}^2_1 + \dot{\hat X}^2_2 \right) 
\end{multline}
where $ \hat X_1$ and $ \hat X_2$ are the (hydrodynamically-rescaled) positions of the tags that diffuse, $c(\rho_1,\rho_2)=c_{\mathrm {eq}}(\rho_1)+c_{\mathrm {eq}}(\rho_2)$  and $\rho_{1,2}=\rho(X_{1,2})$; also $D_t$ is the tag diffusion constant, and the dots indicate time-derivatives (with respect to $\hat{t}$).  The tags can diffuse either by particle diffusion or by transfer of the tag between particles -- for simplicity we take $D_t$ as a simple constant, independent of density.

\begin{figure}
	\centering
	\includegraphics[width=1\linewidth]{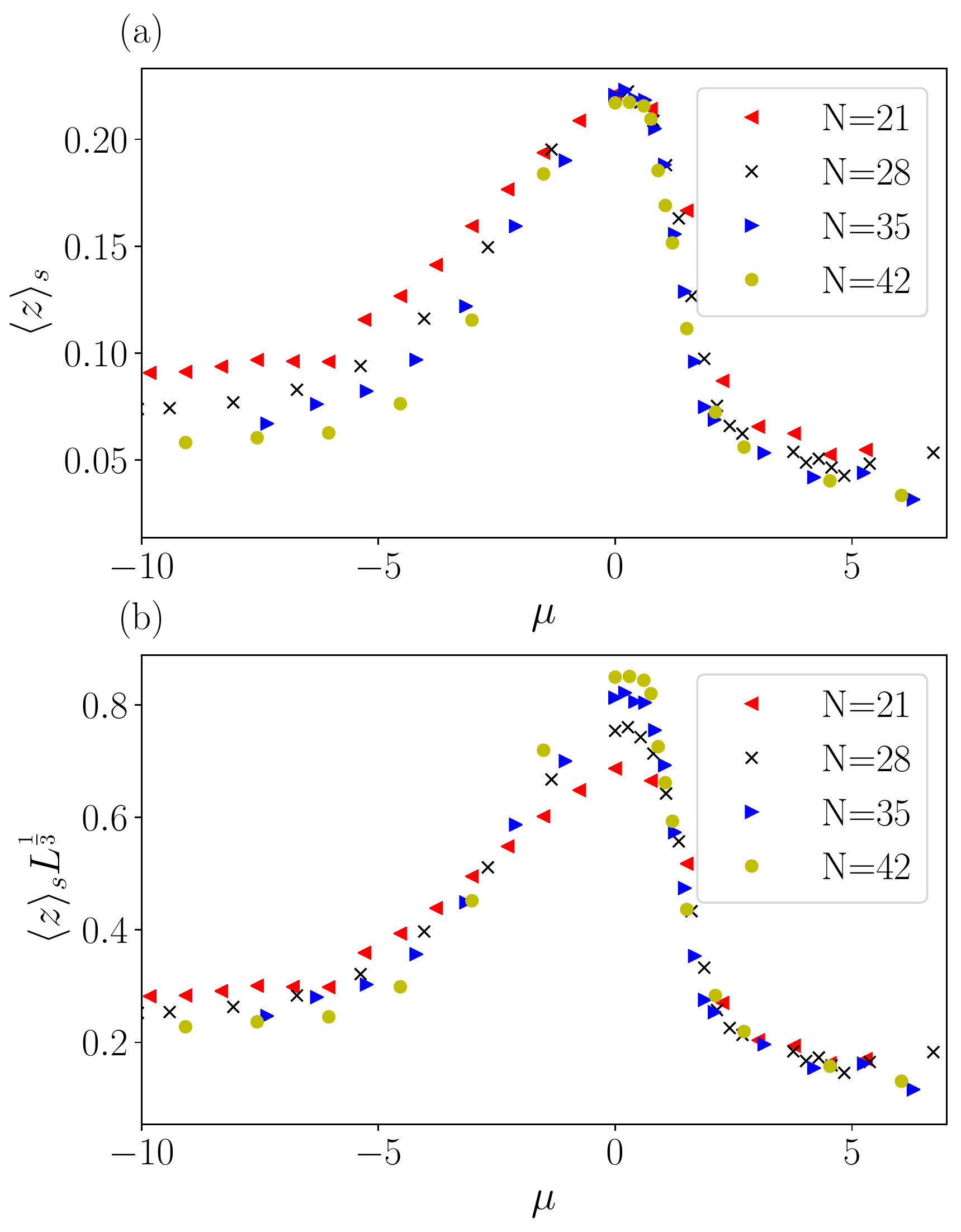}
	\caption{Numerical results for the average distance between two tagged particles, $z=\lvert\hat{x}_1-\hat{x}_2\rvert$ where the system is periodic with size 1 and we use the minimal image convention for the distance, so $z\in[0,0.5]$.  Results are obtained by simulation of the particle model and are shown at different levels of bias $\mu$.  The two panels show a transition between two regimes, where the separation is order $L$ (so $z=O(1)$ at small bias) or of order $L^{2/3}$ (so $z=O(L^{-1/3}$) at large bias). We take $\gammaobs=0.07$ and $\rho_0=7/3$.
	}
	\label{fig:Glike2Part}
\end{figure}

 In contrast to the one-particle case, the tagged particle motion cannot be neglected, so obtaining the dynamical free energy requires integration over the trajectories for $\rho,j$ and also the tag trajectories.  As in the one-particle case, the density and current integrals can be performed by the saddle-point method -- we assume for simplicity that the density and current do not depend on time (consistent with Fig.~\ref{fig:Snapshot1D2Part}).  This leads to a path integral for the tag positions
$
 {\cal Z}_s = \int {\cal D}(\hat{X}_1,\hat{X}_2) e^{- S_2}
 $
 with
 \beq
S_2 = \int_0^{\gammaobs}  \left[ \frac{D_0}{2D_t} \left( \dot{\hat X}^2_1 + \dot{\hat X}^2_2 \right) + \LL V_\mu(X_1-X_2) \right] {\rm d}\hat{t}
\label{equ:S2}
 \eeq
 where 
\begin{equation}
V_\mu=\inf_\rho \bigg[\int_0^{1}  \frac{|\hat\nabla\rho|^2}{8\rho} \, \mathrm{d}\hat{x} \,- \frac{\mu}{2a}c(\rho_1,\rho_2)\bigg]
\end{equation} 
 only depends on the separation of the particles $X_1-X_2$, by translational invariance.

The action (\ref{equ:S2}) describes a biased ensemble of trajectories for two tags with positions $X_{1,2}$ that diffuse freely.  
The corresponding biased ensemble is similar to (\ref{sensemble}) with reweighting factor $\exp[-s\int V(X_1(t)-X_2(t)) dt]$.  Instead of analysing this system via the path integral, the dynamical free energy $\psi(s)$ obeys an eigenvalue problem based on the underlying Fokker-Planck equation~\cite{Jack2020ergo}, which in this case is
\beq
-\frac{D_t}{D_0} P''(y)+\LL V_\mu(y)P(y)=\psi_L(\mu) P(y),
\label{equ:eigen-y}
\eeq 
where $y=\hat X_2-\hat X_1$ is the particle separation and $P(y)$ is the eigenvector.  

The periodic boundary conditions mean that the tags can be considered as moving on a circle, and the separation is measured clockwise from particle $1$ to particle $2$, hence $0<y<1$.   The tags cannot pass through each other so $y(t)$ has reflecting boundary conditions at $y=0$ and $y=1$.  Hence (\ref{equ:eigen-y}) is to be solved subject to $P'(0)=0=P'(1)$~\cite{Touchette2020}.  The symmetry of the problem means that the dominant eigenvector has $P(y)=P(1-y)$, it is sufficient to solve on the domain $0<y<\frac12$ with $P'(\frac12)=0$.

\begin{figure}
	\centering
	\includegraphics[width=1\linewidth]{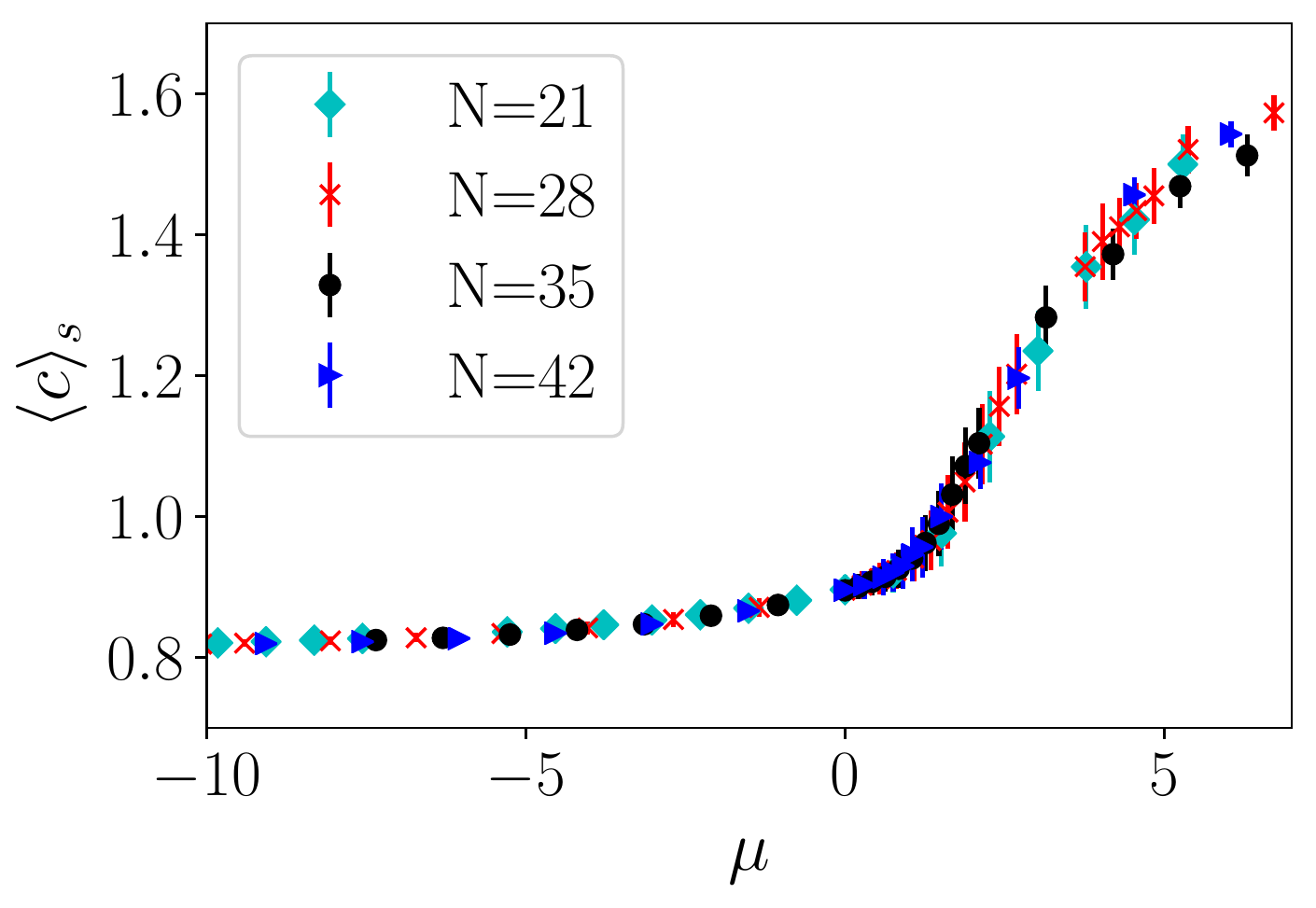}
	\caption{The mean clustering around the two travelling tags for different system sizes, obtained by numerical simulation of the particle model. The clustering is a continuous function of $\mu$ and confirms that in this system we still have a scaling function $\psi_{L}=\frac{1}{L}F(\mu)$, similar to (\ref{equ:psi-L-1d}). 
	Error bars are shown except where they are smaller than symbol sizes. We take $\gammaobs=0.07$ and $\rho_0=7/3$.
	}
	\label{fig:means2part}
\end{figure}

\begin{figure*}
	\centering
	\includegraphics[width=0.8\linewidth]{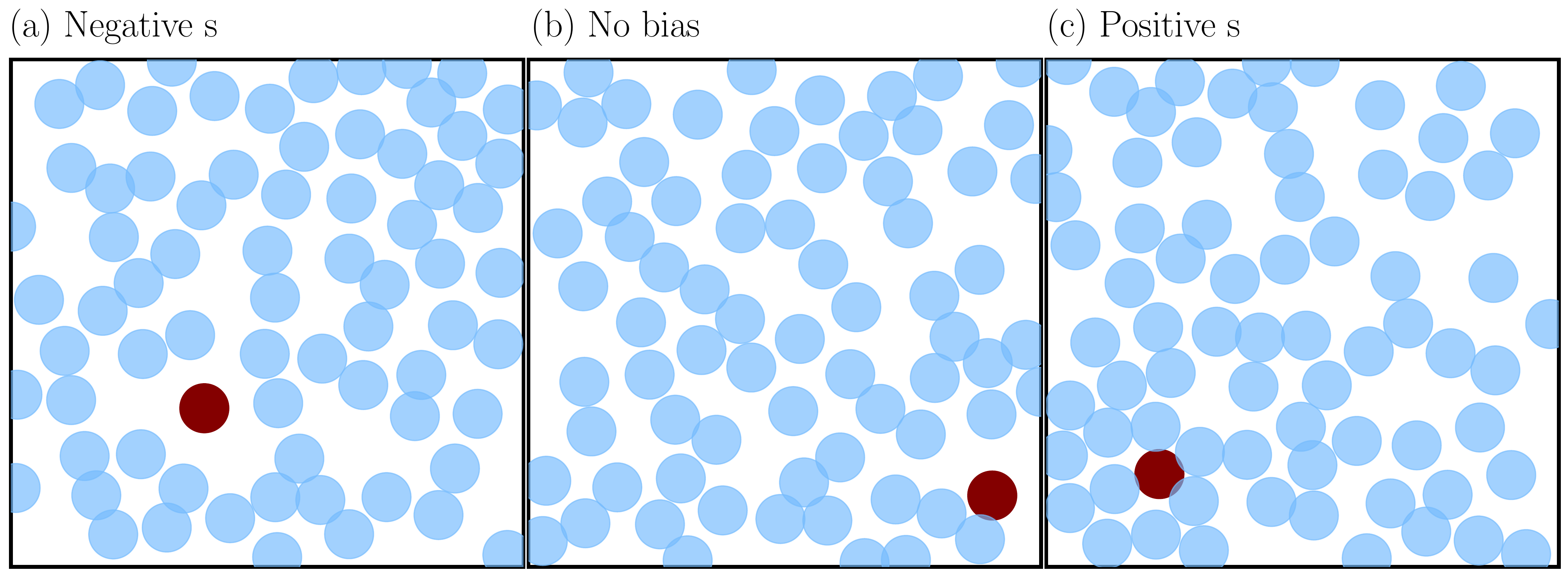}
	\caption{Snapshots of the system under different levels of bias: (a) negative $s$ ($s=-4$) and lower levels of clustering, (b)  no bias at all and (c) positive $s$ ($s=18$) and higher levels of clustering around the selected particle. From a cursory glance it seems that only local effects are caused by this biasing, but we show below that long range effects are also at play. The density is $\rho_0\approx0.47$ throughout this section.}
	\label{fig:2DSnapshots}
\end{figure*}

Equ.~(\ref{equ:eigen-y}) has the form of a  (time-independent) Schr\"odinger equation in which the potential is multiplied by a large parameter $\LL$.  For the unbiased dynamics $s=0$ (so $\mu=0$) then $V_\mu=0$ and the eigenvector $P(y)=1$ is constant.  On the other hand, for $s\neq 0$, the expected behaviour from Fig.~\ref{fig:Snapshot1D2Part} is that the tags tend to localise near $y=0$ or $y=1$.  This effect is independent of the sign of $s$ : the reason is that the two tagged particles can both benefit from the same region of reduced (or increased) density, so both kinds of bias cause them to co-localize.

The function $V_\mu$ could be computed numerically, but the qualitative behaviour of the system can be deduced by the following simple argument: $V_\mu$ has a minimum at $y=0$ and can be approximated as $V_\mu(y) = -V_{\mu}^0 + yV_{\mu}'(0)+O(y^2)$ with $V_{\mu}^0,V_{\mu}'(0)>0$ both dependent on $\mu$.  
As $\LL\to\infty$, the tags localise near $y=0$ (or $y=1$) so this approximate form of $V_\mu$ is sufficient to determine the eigenvector.  In particular Equ.~(\ref{equ:eigen-y}) reduces to the Airy equation and the dominant eigenfunction is the Airy function of the first kind~\cite{vallee2004airy}
\begin{equation}
	P(y)=C_1\;\mathrm{Ai}\bigg[\left(\frac{D_0\LL V_{\mu}'}{D_t}\right)^{1/3}y+ C_2 \bigg],
\end{equation}
where $C_1,C_2$ are constants.  Similar distributions appear in other large-deviation problems, for example~\cite{AiryMajumdarComtet, Bodineau2012jsp, Nemoto2017first}.  For these, $C_2$ is determined by the boundary condition $P'(0)=0$, which requires that the rightmost maximum of the Airy function is at $y=0$~\cite{AiryMajumdarComtet}; then $C_1$ is determined by normalisation.
 One sees from this eigenfunction that if $V_\mu={O}(1)$ (corresponding to $\mu={O}(1)$) then the typical distance $y$  between the tags is $y={O}(L^{-1/3})$.

 \subsection{Discussion and numerical comparison}

The physical content of the preceding analysis is as follows.  From Fig.~\ref{fig::1DSnapshot}, it is clear that biasing the behavior of a single particle causes it to localise inside a macroscopic cluster of particles $(\mu>0)$ or a macroscopic void ($\mu<0$).  On biasing two particles, Fig.~\ref{fig:Snapshot1D2Part} illustrates that the tagged particles tend to co-localise inside a single macroscopic cluster (or void).  An alternative scenario would be that in large systems, the tagged particles get localised in two separate clusters, but this is not observed in practice. Our theoretical analysis shows that  co-localisation in a single cluster (or void), is expected on taking $L\to\infty$ at any fixed value of $\mu$ (except for $\mu=0$).
 
 
 We now compare this theory with numerical results.  Fig. \ref{fig:Glike2Part}(a) shows the behavior of the tag separation $z=|\hat x_1 - \hat x_2|$.  In contrast to $y$, this distance is measured in the minimal image convention so $0\leq z \leq (1/2)$.  For uncorrelated tag positions then $\langle z \rangle = (1/4)$, which is the case for $\mu=0$.  Both positive and negative biasing fields lead to small particle separations, consistent with the theory.
 
 Fig. \ref{fig:Glike2Part}(b) shows that the small distance between the particles scales as $\langle z \rangle_s = {O}( L^{-1/3})$, which is also consistent with theoretical predictions.  This corresponds to a separation of order $L^{2/3}$ when measured in the original co-ordinates (before hydrodynamic rescaling), see also~\cite{Bodineau2012jsp}.  
  Finally, Fig.~\ref{fig:means2part} shows the activity in this biased ensemble, which again shows a scaling form $\langle c \rangle_s = f(\mu)$, consistent with the theory. 
 
\section{Biasing one particle in two dimensions}\label{sec:Results2D}

We now consider a system in $d=2$, using a biased ensemble as in (\ref{sensemble}).  Configurations from representative trajectories are shown in Fig. \ref{fig:2DSnapshots}, which illustrate either increased or reduced clustering around the tagged particle.  In contrast to Fig.~\ref{fig::1DSnapshot}, it is not clear whether the response to the bias is localised around the probe, or if it might be long-ranged.  We will show that in fact the response spans the whole system.

\subsection{Typical fluctuations}

As in one dimension, the role of hydrodynamic effects is already apparent at the level of the variance of $c[\vec{x}]$, via the scaling of $\chi_L(0)$ with $L$.
The density $\rho$ is defined by generalising  (\ref{equ:def-rho-field}) to $d=2$.
The result (\ref{equ:chi-fast-slow}) is still applicable in $d=2$: assuming that the slow contribution dominates in large systems leads to
\beq
\chi_L(0) \approx c'_{\rm eq}(\rho_0)^2 \int_{-\infty}^{\infty} \left\langle \delta\rho(\vec{x}_p(t),t) \delta\rho(\vec{x}_p(0),0) \right\rangle {\rm d}t
\label{equ:chiL-2d}
\eeq
where the approximate equality is accurate for large $L$.  (The neglect of the fast contribution will be discussed below.)  
We 
introduce the Fourier transform of the density
\begin{equation}
\tilde\rho_{\vec{q}}(t) = {L^{-d/2}} \int \rho(\vec{x},t) \exp\left( - {\rm i}\mathbf{q}\cdot\mathbf{x}\right)  {\rm d}^d\vec{x}
\label{equ:fourier2D}
\end{equation} 
with $\bm{q} = \frac{2\pi}{L}(m,n)$ where $n,m$ are integers.  Then the hydrodynamic scaling behavior of (\ref{equ:chiL-2d}) can be obtained by evaluating the correlation function with $\vec{x}_p(t)=\vec{x}_p(0)$, which yields 
\beq  \label{equ:chi-correl-2d}
\chi_L(0) \approx \frac{c'_{\rm eq}(\rho_0)^2 }{L^2} \int_{-\infty}^{\infty} \sum_\mathbf{q}  \left\langle\tilde\rho_\mathbf{q}(t)\tilde\rho_{-\mathbf{q}}(0)
 \right\rangle {\rm d}t
\eeq
The sum over wavevectors is restricted to $|\vec{q}|<Q$ with $Q\approx 2\pi/\Omega$, because of (\ref{equ:def-rho-field}).

At the level of macroscopic fluctuation theory (or fluctuating hydrodynamics) one has
\begin{equation}\label{eqn:MacroscopicFourierDensityCorr}
\langle\tilde\rho_\mathbf{q}(t)\tilde\rho_{-\mathbf{q}}(0)\rangle = ({\sigma}/{D}) {\rm e}^{-D|\vec{q}|^2 |t|}.
\end{equation}
In contrast to the one-dimensional case, the functions $\sigma(\rho)$ and $D(\rho)$ are not known.  
These quantities depend on microscopic details of the system, including the specific choice of interaction potential between the particles (which we take here as a WCA potential).  However, an important feature of MFT is that it makes generic predictions, independent of the detailed dependence of $\sigma,D$ on $\rho$.  For example, (\ref{equ:chi-correl-2d},\ref{eqn:MacroscopicFourierDensityCorr}) yield
\begin{equation}
	\chi_L(0) \approx \frac{1}{L^2}\sum_{\mathbf{q}}^{\mathbf{Q}}\frac{c_{\mathrm{eq}}'(\rho_0)^2\sigma(\rho_0)}{D(\rho_0)^2q^2} \; .
	\label{equ:chiL-2d-sum}
\end{equation}
The summation may be approximated by an integral, see \cite{keta2020collective} for a detailed discussion (specifically Eq. (E14) of that work). The result is
\begin{equation}\label{equa::logLScale}
		\chi_L(0)\approx \frac{1}{4\pi}\frac{c_{\mathrm{eq}}'(\rho_0)^2\sigma(\rho_0)}{D(\rho_0)^2}\ln L + \chi_\infty,
\end{equation}
where $\chi_\infty=O(1)$.  The key result is that the hydrodynamic modes lead to a logarithmic divergence of $\chi_L$ -- this is the dominant contribution in large systems.  The neglect of fast terms in (\ref{equ:chiL-2d}) is justified a posteriori since these would contribute to $\chi_\infty$ but they do not affect the dominant (diverging) term.
The values of $\chi_\infty, D(\rho_0),\sigma(\rho_0)$ etc. depend on microscopic details of the system, but the scaling of $\chi_L(0)$ with $\log L$ is a robust MFT prediction, independent of microscopic details.

\begin{figure}
	\centering
	\includegraphics[width=1\linewidth]{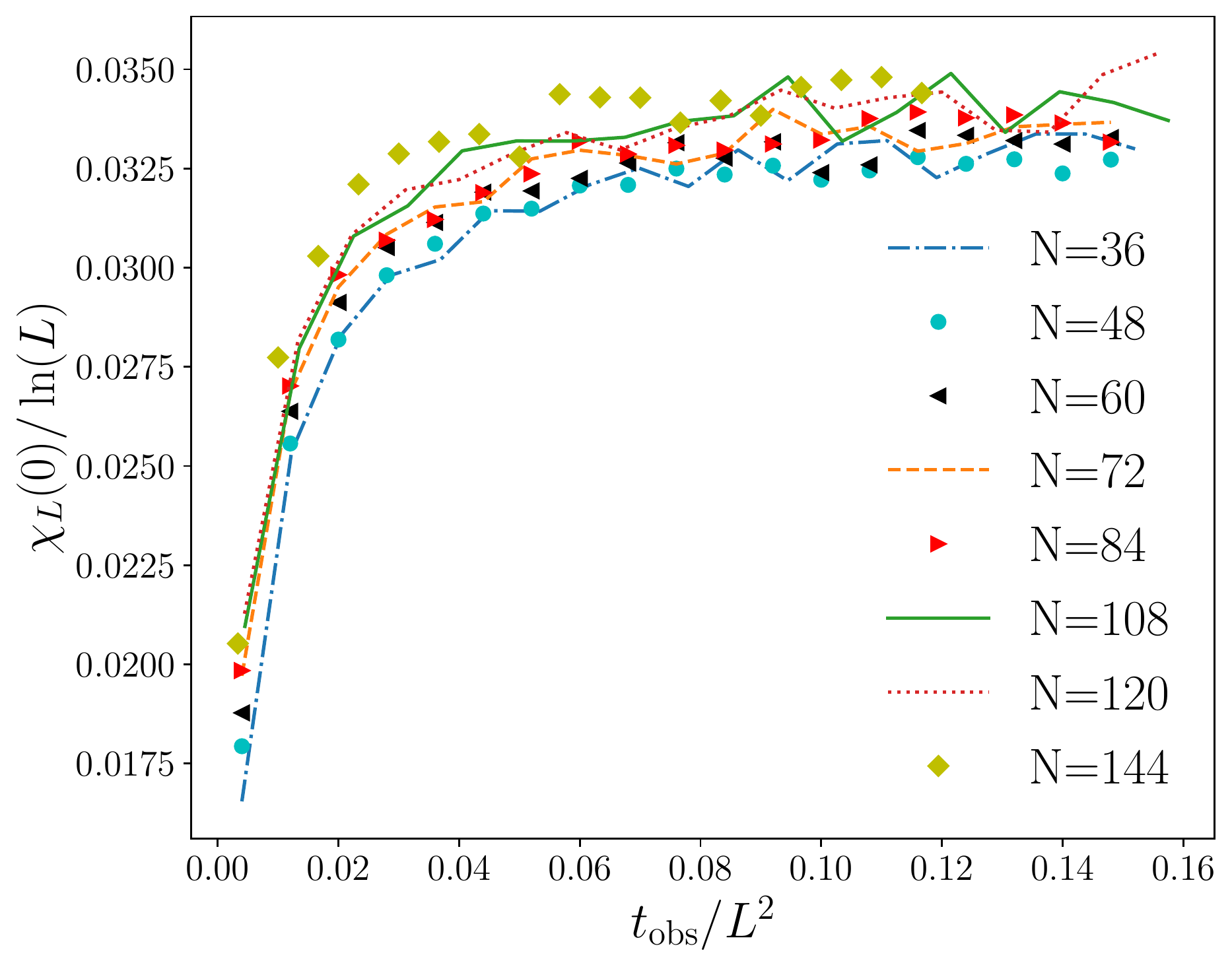}
	\caption{Variance of the clustering in 2D divided by $\ln L$, obtained by numerical simulation of the particle model.  This demonstrates the scaling predicted in \refequa{equa::logLScale}. We apply a diffusive scaling to the observation time to observe a collapse of the curves. The density is $\rho_0\approx0.47$ }
	\label{fig:sigmatobs2d}
\end{figure}

Fig. \ref{fig:sigmatobs2d} shows the behaviour of $\chi_L(0,\tobs)$ from numerical simulations.  As $\tobs\to\infty$, there is good evidence for a logarithmic dependence of $\chi_L$ on $L$, which signals that hydrodynamic modes dominate typical fluctuations, just as they do in one dimension.  [It is notable that the analysis of (\ref{equ:chiL-2d}-\ref{equ:chiL-2d-sum}) is easily generalised to $d\geq 3$; one finds that fast contributions then dominate $\chi_L$ in that case.  [More precisely, the sum in \refequa{equ:chiL-2d-sum} scales as $L^{2-d}$ so this hydrodynamic contribution to $\chi_L$ vanishes as $L\to\infty$ for $d\geq3$.  On the other hand, (\ref{equ:chi-fast-slow}) has a contribution of order unity from fast modes, which determines the value $\chi_L$.]

\subsection{Large deviations}

\renewcommand{\vv}[1]{{\bm{#1}}}
We now turn to large deviations. The equivalent expression to \refequa{equ:action-rho-j} in $d$ dimensions (assuming that diffusion and mobility are isotropic) is 
	 \begin{multline}
	\label{equ:action-rho-j-2D}
	S[\rho,j, x_p] = L^d  \int_0^{\gammaobs}  \int_{[0,1]^{d}} \frac{|D_0\hat \jmath + D(\rho)\hat\nabla\rho/2|^2}{2D_0\sigma(\rho)} \, \mathrm{d}^d\hat{\vv{x}} \, \mathrm{d}\hat{t}
	\\- \frac{sL^2}{2D_0}  \int_0^{\gammaobs}  c_{eq}(\rho) \mathrm{d}\hat{t},
\end{multline} where $D_0$ is the diffusion coefficient at $\rho=0$,  also $\hat{\vv{x}}$ is the position vector in $d$ dimensional  space (after hydrodynamic rescaling) and we recall that $\gammaobs= \frac{\tobs}{\tau_L}= \frac{2D_0\tobs}{L^2} \;$.

Comparing the relative sizes of the terms in the first and second lines of (\ref{equ:action-rho-j-2D}), there is a strong dependence on dimensionality.  For $d=1$ then it is necessary to take $s\propto 1/L$ in order that the biasing term (proportional to $s$) has the same $L$-dependence as the first (hydrodynamic) term.  Hence the response to the bias is controlled by the scaling variable $\mu\propto sL$, as defined in (\ref{equ:def-mu}).   For $d=2$ then the biasing term has the same scaling as the hydrodynamic one for $s$ of order unity.  In this case one should expect  such a bias to generate a long-ranged (hydrodynamic) response in the density: this will be verified below.  
{For $d\geq3$ then one would require $s\propto L^{d-2}$ in order for the two contributions in (\ref{equ:action-rho-j-2D}) to have the same scaling.  However, such large biases are outside the scope of the hydrodynamic theory.  [To see this, note that biasing fields of order unity already induce significant responses in the fast modes of the system, leading to changes in the local (microscopic) structure.  For large bias ($s\propto L^{d-2}$), the microscopic structure will be completely different from its equilibrium state, contrary to the assumptions of the hydrodynamic theory.]
In practice, we expect interesting behavior for $d\geq 3$ and bias $s$ of order unity, due to the response of fast modes to the bias.  That is, a local biasing field --  acting on a single probe particle -- should significantly increase the density in the vicinity of that particle. This response will be short-ranged and cannot be computed by the hydrodynamic theory.}

 \begin{figure}
	\centering
	\includegraphics[width=1\linewidth]{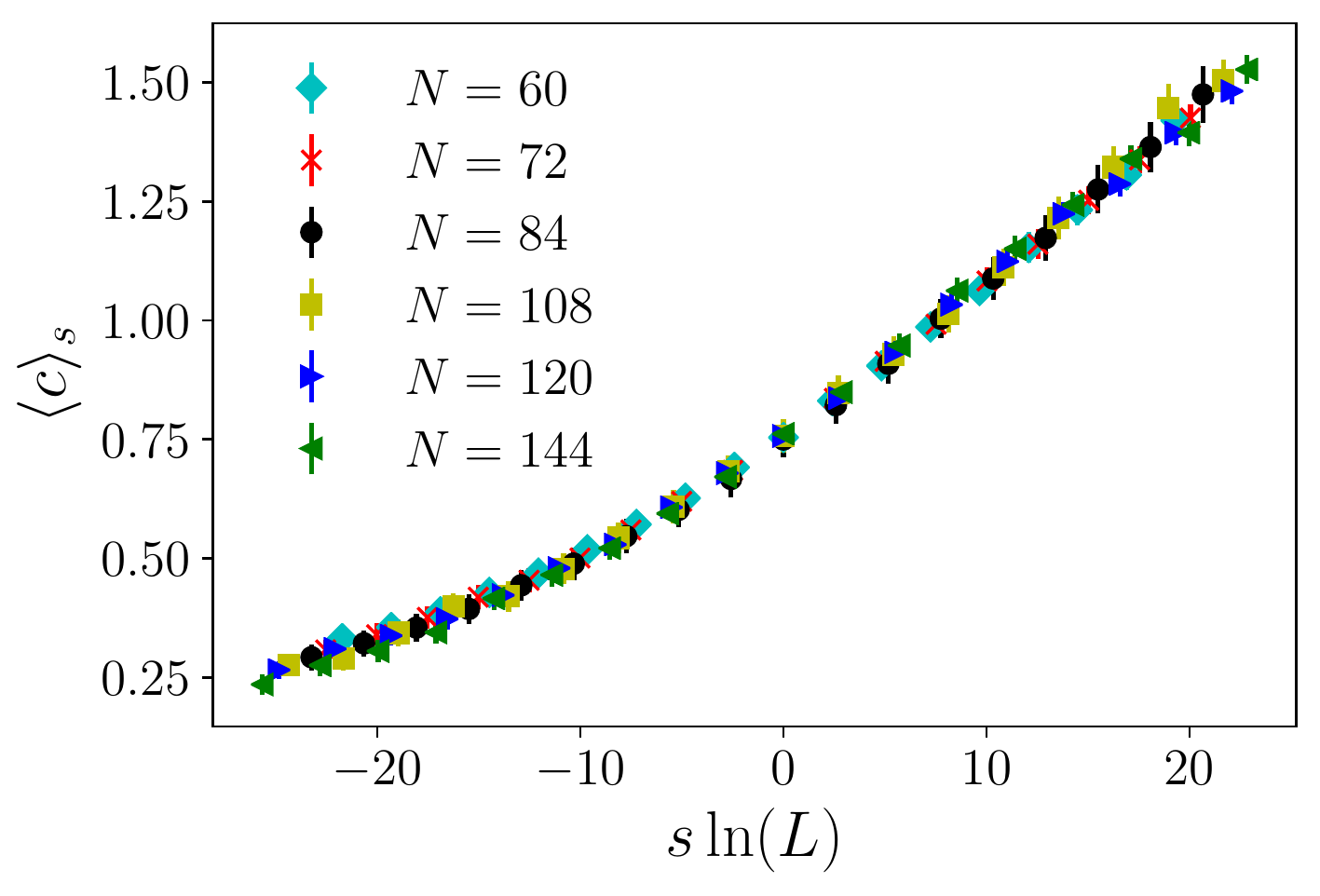}
	\caption{Results for the clustering in 2D, as a function of bias $\mu$, obtained by numerical simulation of the particle model. The data collapse as a function of $s\ln L$ is consistent with $c \sim F(s\ln L)$, which corresponds to (\ref{equ:psi_L_2d}).  Here and for the rest of this section $\gammaobs=0.08$.  }
	\label{fig:means2d1particle}
\end{figure}

 
  Fig.~\ref{fig:means2d1particle} shows the behaviour of $\langle c \rangle_s = \psi_L'(s)$ in this two-dimensional system.  
The logarithmic divergence of $\chi_L$ in (\ref{equa::logLScale}) raises a question as to the analogue of (\ref{equ:psi-L-1d}) in this case.  One possibility that is consistent with $\langle c\rangle_s=O(1)$ as well as a logarithmic divergence of $\chi_L$ is
\beq
\psi_L(s) \approx \frac{1}{\ln L} \Psi(s\ln L) \; .
\label{equ:psi_L_2d}
\eeq
This would suggest that $s\ln L$ is a useful scaling variable, similar to $sL$ in the $1d$ case.   We
do not have any mathematical argument in favour of (\ref{equ:psi_L_2d}) but Fig.~\ref{fig:means2d1particle} shows that data for several system sizes collapse when plotted as a function of this variable. (This is consistent with the scaling of $\chi_{L}$ derived in  \refequa{equa::logLScale} and shown in Fig.~\ref{fig:sigmatobs2d}.).  

Finally, we return to Fig.~\ref{fig:2DSnapshots} and the question of whether the bias on a single particle elicits a system-wide response.  
We first investigate the response to the bias of the Fourier components of the density  
\begin{equation}
	\frac{d}{ds}\langle \tilde{\rho}_{\mathbf{q}}(t) \rangle_s=\big\langle \tilde{\rho}_{\mathbf{q}}(t)c[\mathbf{x}] \big\rangle,
\end{equation}
 where we used \refequa{equ:ave-s}.  Now use the definition of the clustering of a trajectory (\ref{equ:cbar}), split the clustering into its slow and fast contributions (as in \refequa{eqn:cslow+fast}) and neglect the fast contribution.  Performing a Taylor expansion of the slow contribution to the clustering in terms of the Fourier components yields
\begin{equation}
	\frac{d}{ds}\langle \tilde{\rho}_{\mathbf{q}}(t)\rangle=\frac{1}{L^{d/2}}\sum_{\mathbf{q}'} \int_{-\infty}^{\infty} \Big\langle \tilde{\rho}_{\mathbf{q}}(t)  
	c'_{eq}(\rho_0) \tilde{\rho}_{\mathbf{q}'}(t')
	{\rm e}^{i{\mathbf{q}'\cdot\mathbf{x}_p}}
	 \Big \rangle \mathrm{d}t'
\end{equation}
The only nonzero components of the sum have $\bm q'=-\bm q$ since all other Fourier components are uncorrelated and do not contribute. Considering this on the level of macroscopic fluctuations, working near equilibrium and using \refequa{eqn:MacroscopicFourierDensityCorr} this expression reduces to
\begin{equation}\label{eqn::derivativeRhoqs}
	\frac{d}{ds}\langle \tilde{\rho}_{\mathbf{q}}(t)\rangle=  \frac{\sigma(\rho_0) c'_{eq}(\rho_0)\exp(-i{\mathbf{q}\cdot\mathbf{x}_p})}{D(\rho_0)^2q^2L^{d/2}},
\end{equation}
where the mobility and diffusion were approximated by their equilibrium values.

\newcommand{\Sq}{\tilde{S}}

This result is already sufficient to verify the general arguments from above, about conditions for observing macroscopically inhomogeneous systems, as a function of dimension $d$.  To see this, note the normalisation convention in (\ref{equ:fourier2D}), which means that a
homogeneous system with Gaussian fluctuations has (typically) $\rho_{\bm q}$ of order unity, as $L\to\infty$.  For macroscopically inhomogeneous systems $\langle\rho(\hat x)-\rho_0\rangle_s$ should be of order unity, corresponding to  $\rho_{\bm q}$ of order $L^{d/2}$.  (This assumes macroscopic wave vectors, $|\bm{q}|$ of order $L^{-1}$.)  Under these assumptions (\ref{eqn::derivativeRhoqs}) predicts for $d=1$ that the system will be macroscopically inhomogeneous as soon as $s=O(1/L)$.  In $d=2$ the corresponding condition is $s=O(1)$.  For $d\geq 3$ then a bias $s$ of order unity is not sufficient to create a macroscopically inhomogeneous system.

For a comparison with numerics, it is convenient to consider the structure factor, defined as 
\begin{equation}
	 \Sq_s(\vec{q})= \left\langle \tilde{\rho}_{\mathbf{q}}\tilde{\rho}_{-\mathbf{q}} \right\rangle_s \; .
\end{equation} 
 Recalling again (\ref{equ:fourier2D}), this quantity is of order unity in a homogeneous system.  For an inhomogeneous system, one expects $\langle \tilde{\rho}_{\mathbf{q}}\tilde{\rho}_{\mathbf{-q}}\rangle_s \approx |\langle \tilde{\rho}_{\mathbf{q}}\rangle_s|^2 \sim L^d$.  Hence, while snapshots like those of Fig.~\ref{fig:2DSnapshots} do not provide a clear distinction between macroscopic and microscopic (finite) clusters, computation of $\Sq_s(\vec q)$ can be combined with a finite-size scaling analysis.

To this end,we  focus on the structure factor associated with the smallest allowed wavector, which has modulus $2\pi/L$.  From \refequa{eqn::derivativeRhoqs}, the leading contribution (in $s$) to the structure factor at this wavevector is predicted as follows: For $d=1$ then $\Sq_s(\vec{q}_1) \propto \mu^2 L$, corresponding to macroscopic inhomogeneity as soon as $\mu=O(1)$.  For $d=2$ the corresponding result from \refequa{eqn::derivativeRhoqs} is
\beq
\Sq_s(\vec{q}_1)-\Sq_0(\vec{q}_1) \approx \frac{s^2 L^2}{4\pi^2}  \left[\frac{\sigma(\rho_0) c'_{eq}(\rho_0) }{D(\rho_0)^2}\right]^2
\label{equ:Ss-dim2}
\eeq
 which predicts macroscopic inhomogeneity as soon as $s=O(1)$.  [For $d>4$ the corresponding contribution vanishes as $L\to\infty$ so hydrodynamic modes are irrelevant at this limit.  
For $d=3$ then $\Sq_s(\vec{q}_1) \propto s^2 L$ which indicates that the system is macroscopically homogeneous, but has anomalous density fluctuations.  We restrict our discussion here to $d=1$ and $d=2$, but higher dimensions such as $d=3$ could be of interest in future work.]

\begin{figure}
	\centering
	\includegraphics[width=1\linewidth]{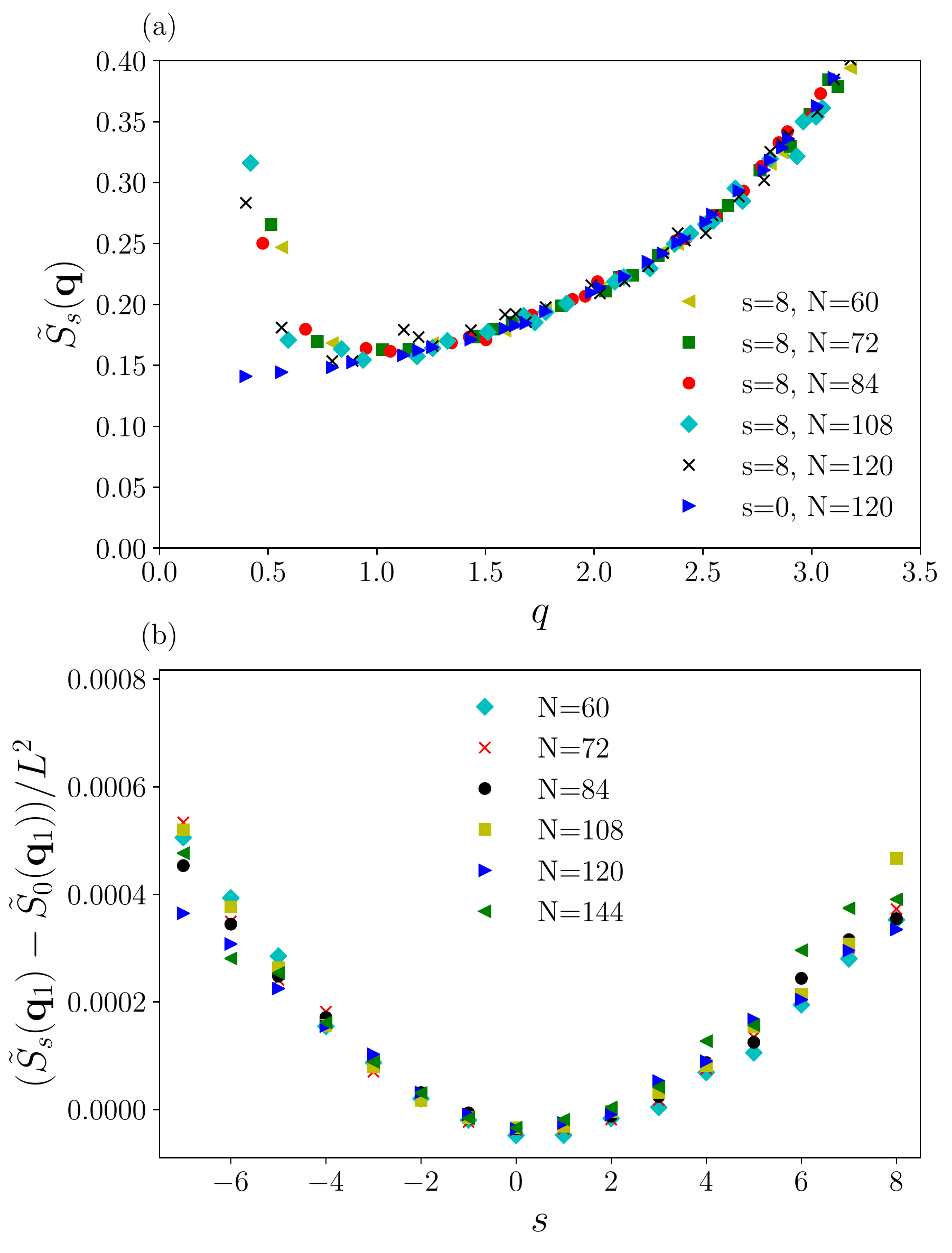}
	\caption{(a) Plot of the structure factor at $s=0$ and $s=8$ for different system sizes in 2D.  At $s=8$ the first few Fourier components are increased. This indicates that the response to the bias is long-ranged {and $q$-dependent}. 
		(b) Plot of the structure factor of the first Fourier component and its dependence on the bias $s$.  The data collapses when plotted in this way, {consistent with (\ref{equ:Ss-dim2}).
		(All results in this Figure were obtained by numerical simulation of the particle model.)
	}
}
	\label{fig:scombined}
\end{figure}

Similar to the logarithmic scaling in (\ref{equa::logLScale}), MFT leads to a robust prediction (\ref{equ:Ss-dim2}), for the dependence of $\Sq_s(\vec q)$ on  $s$ and $L$.  This prediction  is independent of microscopic details such as the interaction potential between particles; it is also straightforward to test numerically.
From TPS simulations, we estimate the Fourier transform of the density as $ \tilde{\rho}_{\mathbf{q}} = L^{-d/2} \sum_j e^{-{\rm i}\vec{q}\cdot\vec{x}_j}$. 
Fig.~\ref{fig:scombined}(a), shows that in a two dimensional system $\Sq_s(\bm{q})$ responds most strongly to $s$ at the smallest wavevectors.  [A local response would mean that $\Sq_{s}(\mathbf{q})$ responds only for wavectors $q=O(1)$, but here the response is large for very small wave vectors $q=O(L^{-1})$.] 
Fig.~\ref{fig:scombined}(b) shows the response at the smallest allowed wavevector.  The result is consistent with (\ref{equ:Ss-dim2}), showing that the numerical and analytical approaches are once again consistent with one another.

\section{Conclusion}\label{sec::Summary}

We have discussed large deviations of clustering around tagged particles, and associated collective behaviour.
All the systems considered show collective responses to the the bias, on length scales that are comparable to the system size. 

In $d=1$ the response depends on a scaling variable $\mu\propto sL$; for $d=2$ the corresponding variable is $s\ln L$.  We have also found that the dynamical free energy has the scaling form (\ref{equ:psi-L-1d}) in one dimension, and our corresponding conjecture in two dimensions is (\ref{equ:psi_L_2d}).
The systems can be analysed through the lens of MFT under the assumption that \refequa{equ:psi-var-1d} captures the relevant fluctuations.
From these we obtain information about the density profile \refequa{equ:ABC-rho} and \refequa{equ:ABC-rho-neg} and the associated clustering around the single particle which we plot in figures \ref{fig:maxdensitysimulations} and \ref{fig::1DMeans} respectively.  Combining these results with Fig.~\ref{fig:1DVariance}, we observe long range correlations in this system, which are caused by hydrodynamic effects. 

 One can understand these as local correlations in the rescaled MFT system with $\hat{x}\in[0,1]$ and $\hat{t}\in[0,1]$.  Hence we can explain these long-range correlations, even if they are not expected from the analogy between dynamical large deviations and thermodynamics.  A key strength of this approach is that MFT provides robust predictions for the scaling of different observable quantities with system size $L$ and bias $s$.  That is, the results do not depend on microscopic details of the system, such as the interaction potential.  This can be appreciated by the fact that the scaling predictions of MFT are accurate both for hard particles in $1d$ and for WCA particles in $2d$.  Quantitative predictions depend on knowledge of the functions $D(\rho)$ and $\sigma(\rho)$, and on a joint limit of $L,\gammaobs\to\infty$.  In this work, the functions $D,\sigma$ are known in $d=1$ but not in $d=2$.  The range of $L,\gammaobs$ that can be considered numerically also hinders quantitative matches between theory and numerics.  Still, the numerical results are consistent with MFT scaling, and the predictions in $1d$ are semi-quantitative.

Compared with previous work where ensembles were biased by quantities that depend on all particles at once~\cite{lecomte_inactive_2012,Nemoto2019,YongJooDynamicalSymmetryBreaking,Jack2015,Dolezal_2019}, 
we bias here one or two particles.  This explicitly breaks the translational symmetry of the system.  Hence this symmetry cannot be spontaneously broken, in contrast to~\cite{lecomte_inactive_2012,Dolezal_2019}.
 Comparing with~\cite{Cagnetta2017,Grandpre2018}, our results accentuate that very large time scales (of order $L^2$) should be considered when analysing large deviations in systems with hydrodynamic modes, even for single-particle quantities.  The coupling of these modes to large deviation events is important as $\tobs\to\infty$, but this can easily be missed, even if the values used for $\tobs$ are larger than all microscopic time scales: values of order $L^2$ are required.
Consistent with other studies of large deviations of local quantities~\cite{Imparato2009,Krapivsky2014,Dhar_2014,GarrahanPretransition, BrownianParticleBurgers, Cagnetta2017, Grandpre2018}, our results illustrate a rich phenomenology that appears in biased ensembles, including dynamical free energies that have scaling forms for large $L$.  In particular, it is not safe to assume that local biases result in local responses, in these ensembles of trajectories.

Finally, we note that since previous work considered biases acting on all particles~\cite{Dolezal_2019} and we considered here biases on one or two particles, a natural question would be what happens for a bias on an intermediate number.  For example, what about tagging a finite fraction of particles?  We have not investigated such situations, but we offer some preliminary remarks.  First, biasing a finite fraction of particles would change the scaling (with $L$) of the bias term in (\ref{equ:action-rho-j-2D}), leading to a situation similar to~\cite{thompson_dynamical_2015,Dolezal_2019}, where all particles were biased. In that sense, biasing a finite fraction of particles is more similar to biasing all of them, and less similar to the situation considered here.  Second, the nature of any clustering transitions would be different when biasing a finite fraction of particles -- the most appropriate analogy might be that of de-mixing of tagged/untagged species, as distinct from clustering of a single species.  Such effects might be usefully investigated in future work.

The data underlying this publication will be available shortly after
publication at https://doi.org/10.17863/CAM.69871 .

\begin{acknowledgments}
We thank Juan Garrahan, Fr\'ed\'eric van Wijland, Tal Agranov, and Yann Keta for helpful discussions.
JD was supported by a studentship from the EPSRC, reference EP/N509620/1.
\end{acknowledgments}
 
\bibliographystyle{unsrt}
\bibliography{dev,Bibliography}

\end{document}